\documentclass[10pt,a4paper]{article}

\usepackage[a4paper, left=2.5cm, right=2.5cm]{geometry}
\newcommand{\sch}{Schr\"odinger}
\usepackage{amsmath,amssymb}
\usepackage{amsfonts}
\usepackage[dvipsnames]{xcolor}
\usepackage{graphicx}
\usepackage{dcolumn}
\usepackage{comment}
\usepackage{authblk}

\usepackage{epstopdf}
\usepackage{color}

\usepackage{amsthm}
\usepackage{fontenc}

\usepackage{soul}

\usepackage{bm}
\RequirePackage{slashed}\usepackage{cancel}
\usepackage[normalem]{ulem}

\def\be{\begin{eqnarray} &&} 
 
\def\ee{\end{eqnarray}} 
\usepackage{array,booktabs}
\newcolumntype{C}{>{$\displaystyle}c<{$}}

\def\be{\begin{eqnarray} &&} 
 
\def\ee{\end{eqnarray}}

\newcommand{\bfi}{\begin{figure}}
\newcommand{\efi}{\end{figure}}

\usepackage{multirow}
\usepackage{hhline}
\usepackage{tikz}
\usepackage{makecell}
\usepackage{comment}

\definecolor{tcA}{rgb}{0,0,0}
\usepackage{array}
\newcolumntype{P}[1]{>{\centering\arraybackslash}p{#1}}

\usepackage{hyperref}
\hypersetup{colorlinks=true, urlcolor=blue, linkcolor=blue, citecolor=blue}
\usepackage{multicol}

\usepackage{xcolor}

\usepackage{titlesec}

\titleformat{\section}{\large\bfseries}{\thesection}{1em}{}
\titleformat{\subsection}{\normalsize\bfseries}{\thesubsection}{1em}{}

\begin{document} 
\title{ Pion and $\eta$ meson gravitational form factors from GPDs at non-zero skewness and distribution amplitudes  within the Graviton Soft-Wall model}

\date{}
\newcommand{\abstractText}{\noindent
We present the calculations of the pion and $\eta$ meson gravitational form factors (GFFs) from the
evaluation of generalized parton distribution functions at non-zero skewness within the graviton soft-wall (GSW) model of hadron structure. Results for the pion have been successfully  compared with lattice data for both the $A$ and $D$ GFFs. From these quantities,  
 the mechanical properties of the pion, i.e.,  the pressure, shear-force, normal and tangential force distributions have been also investigated.
Finally, we study also the $\eta$ meson GFFs to  predict and observe differences between the pion and $\eta$ structures in the corresponding mechanical properties within the GSW model.

}

\author[1]{Filippo Fornetti}

\author[2]{Matteo Rinaldi\thanks{Corresponding author email: matteo.rinaldi@pg.infn.it}}


\author[1,2]{Sergio Scopetta \thanks{S. Scopetta is presently Science Counsellor at the Italian Embassy in Spain, calle Lagasca 98, 28006 Madrid, Spain. sergio.scopetta@esteri.it}}

\affil[1]{\rule[25pt]{0pt}{0pt} Dipartimento di Fisica e Geologia, Universit\`a degli Studi di Perugia and INFN Sezione di Perugia, Italy.}

\affil[2]{\rule[25pt]{0pt}{0pt} Istituto Nazionale di Fisica Nucleare, Sezione di Perugia, Via A. Pascoli, Perugia, Italy.}



    \maketitle
    \begin{abstract}
     \abstractText
    \end{abstract}

\section{Introduction}
In hadron Physics, despite the successes  of  the Quantum Chromodynamics (QCD)  predictions when compared with the data, several properties of baryons and mesons are still unknown due to the high non linearity of the theory, which prevents their evaluations. In this study, we are interested in the  
 energy-momentum tensor
(EMT) matrix elements which encodes, for example, information on the mass and the spin origin in hadrons \cite{Polyakov:1999gs,Kobzarev:1962wt,Pagels:1966zza}. In particular, we focus on the so called gravitational form factors (GFFs) which parametrize the Lorentz structure of the matrix elements of the EMT. For example, the GFF $D$ can be ascribed to the 
variation of the spatial components of the  metric and, hence to  deformations. Therefore, this quantity 
 can be studied to access 
the force distributions inside hadrons and the corresponding “mechanical properties”.
In principle, GFFs should be measured in gravitational scattering processes, where hadrons interact with gravitons and the corresponding cross-section would depend on the parametrization of the EMT. However, for the moment being such a scenario is beyond the present experimental feasibility. Nevertheless, GFFs can be linked to moments of generalized parton distribution  \cite{Ji:1996ek,Diehl:2003ny} which can be accessed through exclusive processes such as the deeply virtual Compton scattering.
From a theoretical point of view, let us mention some attempts to estimate the proton and pion GFFs within lattice QCD approaches  \cite{Hackett:2023nkr,Shanahan:2018pib}. In this framework, we are interested in the  pion  and $\eta$ meson GFFs in order to eventually emphasize differences in their mechanical properties due to the distinct underline dynamics. To this aim we make use of a quark model based on the AdS/QCD Light-Front approach \cite{Brodsky:2007hb,deTeramond:2005su,deTeramond:2008ht,Gutsche:2014yea,Chakrabarti:2015lba}. We remind that originally
 the holographic principle relies in a correspondence between a five dimensional classical theory with an AdS metric
and a supersymmetric conformal quantum field theory with $N_C \to \infty$. Despite 
 the latter is
different from QCD, the gravity sector of the theory can be properly modified for phenomenological purposes,  the so called bottom-up approach \cite{Polchinski:2000uf,Abidin:2008sb}.
In this framework, models differ in how they implement 
the conformal invariance breaking. 
In this analysis we make use of the 
graviton soft-wall model (GSW) \cite{Rinaldi:2017wdn,Rinaldi:2018yhf,Rinaldi:2020ssz,Rinaldi:2021dxh,Rinaldi:2022dyh,Rinaldi:2024fgx}  based on the 
 holographic Soft-Wall (SW) model. Within the latter,  
a  dilaton field is introduced in the gravitational action to softly break conformal invariance \cite{Karch:2006pv}. However, in order to predict the linear spectrum of the scalar glueballs
the GSW model has been introduced by modifying the original AdS metric.
 Later on, the approach has been successfully tested to reproduce the spectra of: mesons, high spin glueballs, and  hybrids \cite{Rinaldi:2024fgx}. Relevant for this study is the good agreement of the predicted pion properties provided by the GSW model with present data.    In general, these approaches have been also used to evaluate the
proton and pion GFFs \cite{Chakrabarti:2015lba,Li:2023izn,Abidin:2008hn,Brodsky:2008pf,Hippelainen:2026izm,Nascimento:2025smf}. 
Here, the pion GFFs, obtained within the GSW model, have been also successfully compared with lattice data. Furthermore,
 the $\eta$ meson GFFs have been calculated and investigated.
The last step is essential to address possible differences between the mechanical properties of
these two pseudo-scalar systems.
Finally, the pion and $\eta$ GFFs have been also evaluated at high momenta, in the perturbative regime, starting from the corresponding distribution amplitudes (DAs) by using the procedure of Refs. \cite{Gao:2022vyh,Tong:2021ctu,Tong:2022zax}. Also in this case, results agree with those of Ref. \cite{Tong:2022zax}, where lattice  DA have been used to evaluate the corresponding GFFs.
The manuscript is organized as follows: \textit{i)} in Sect. 2 we formally introduce the gravitational form factors and related quantities that will be evaluated; \textit{ii)} in Sect. 3 the pion and $\eta$ wave-functions obtained within the graviton soft-wall model is presented; \textit{iii)} in Sect. 4 the meson GPDs in temrs of the corresponding hadron wave-function are written; \textit{iv)} in Sect. 5 numerical results are shown and discussed;\textit{v)} in Sect. 6 the GFFs obtained from the corresponding meson distribution amplitudes are calculated and compared with lattice results.


\section{Gravitational Form Factors and GPDs}
In this section we introduce the   gravitational form factors  defined from the Lorentz decomposition of the Energy-Momentum tensor matrix element. This quantity can be defined as:

\begin{align}
    \hat T_{\mu \nu} = \sum_a \hat T_{\mu \nu}^a;
\end{align}
being $\hat T_{\mu \nu}^a$ the contribution of the parton of flavor $a=q,g...$ to the total energy momentum tensor.
In this analysis  we make use of the conventions of Ref. \cite{Polyakov:2018zvc} for spinless targets, like the pion and the $\eta$, with initial (final) momentum $p$ $(p')$. In this framework, one has:

\begin{align}
    \langle p' | \hat T^a_{\mu \nu}(0)|p\rangle = 2 P_\mu P_\nu A_a(t)
+ \frac{1}{2}(\Delta_\mu \Delta_\nu -g_{\mu \nu}\Delta^2)D_a(t)+m^2 \bar c^a(t),
\end{align}
where $P= (p+p')/2$, the space-like vector $\Delta = p'-p$,  and $t = \Delta^2$. Finally, $A_a(t)$, $D^a(t)$ and $\bar c(t)$ are the GFFs with the constraint $\sum_a \bar c_a(t)=0$ from the conservation of the EMT, i.e. $\partial^\mu \hat T_{\mu \nu}=0$. In addition, from translational invariance  and momentum conservation:
\begin{align}
    A(0) = \sum_a A_a(0)=1.
\end{align}
In  the case of the $D(t)$ GFF, the quantity

\begin{align}
D(0) =D= \sum_a D_a(0)
\end{align}
is not generically constrained.
Nevertheless, for the pion in the chiral limit, 
one has \cite{Polyakov:1999gs,Hudson:2017xug,Donoghue:1991qv}:

\begin{align}
    D^\pi(0) = -1.
\end{align}
We stress that GFFs
 encode fundamental information on the partonic inner structure of hadrons, for example, 
the $D(t)$ GFF  is intimately related to pressure and shear forces due to the strong interaction. 
Therefore, future  extraction of GPDs and GFFs will open new windows on the mechanical properties of hadrons. 
 In fact,
the gravitational form factors can be used to directly evaluate all components of the EMT which provide, e.g.,  the energy density of the hadron \cite{Polyakov:2018zvc}:

\begin{align}
    T_{00}(r) &= 2 m^2 \int \dfrac{d^3 \Delta}{2 (2 \pi)^3 E} e^{-i \vec r \cdot \vec \Delta} \left\{A(t)- \dfrac{t}{4 m^2}\Big[A(t)+D(t) \Big] \right\}~,
\end{align}
while the pressure and the shear forces, obtained from the spatial components of the EMT, can be related to the $D(t)$ GFF:

\begin{align}
\label{Eq:p}
    p(r) &= \dfrac{1}{3} \int \dfrac{d^3 \Delta}{2 (2 \pi)^3 E} e^{-i \vec r \cdot \vec \Delta}~t D(t)
    \\
        s(r) &= \dfrac{3}{4} \int \dfrac{d^3 \Delta}{2 (2 \pi)^3 E} P_2(\cos \theta) e^{-i \vec r \cdot \vec \Delta}~t D(t)
        \label{Eq:s}
\end{align}
where  $E= \sqrt{m^2+\vec \Delta^2/4}$. 
Since the pressure distribution has been defined in terms of  $D(t)$, one can rely the  Von Laue condition, i.e. the balance of forces inside  
 \cite{Polyakov:2018zvc,Laue:1911lrk}:

\begin{align}
    \int d \vec r~ p(r) =0.
    \label{Eq:condition}
\end{align}
to a constraint for the $D(t)$ GFF (cfr.  Eq. \eqref{Eq:p}):

\begin{align}
    \underset{t \to 0}{lim} ~t D(t)=0.
\end{align}
 
As already stated, despite the crucial information encoded in the GFFs, their extraction from processes involving gravitational interaction is prohibitive. Nevertheless, it has been proved that 
 GFFs  are related   to moments of GPDs \cite{Ji:1996ek,Diehl:2003ny}. 
We exploit this connection to numerically extract the pseudo-scalar meson GFFs from GPDs computed using the Light-Front (LF) wave function (wf) of the GSW model.
 In this case, the second moment of the only GPD $H$ reads as follows:
\begin{align}
    \int_{-1}^1 dx~x H^a(x,\xi,t)=A_a(t)+\xi^2 D_a(t)~,
    \label{Eq:poly}
\end{align}
where, $x$ is the mean longitudinal momentum fraction carried by a parton in the hadron between the initial and final state and $\xi$, the so-called skewness, is the longitudinal momentum transferred fraction w.r.t. the mean hadron momentum. 
We remind that the polynomial structure of 
Eq. (\ref{Eq:poly}) is due to Lorentz covariance \cite{Diehl:2003ny,Ji:1998pc}. The above relation is therefore a key pass to access the GFFs from a  process whose cross-section is used to extract  GPDs. 
Notably, in some cases, GFFs can be evaluated directly within holographic models from the gravitational action in AdS space\cite{Li:2023izn,Liu:2025qja,Liu:2025din,Fujii:2024rqd,Hippelainen:2026izm,Cao:2026jzm}. 
 However, since calculating different parton distribution functions is fundamental for testing models, we evaluate GFFs directly from GPDs in the present analysis.
 It is worth emphasizing that, for the proton, AdS/QCD models are suitable only for the evaluation of  GPDs  at zero skewness from the corresponding LF wave-functions (see, e.g., Refs. \cite{deTeramond:2018ecg}). This restriction stems from the fact that these models do not yield the complete LF hadronic wave-function in terms of all parton momenta, a prerequisite for computing GPDs at non-zero skewness (see, e.g., Ref. \cite{Rinaldi:2017roc}).  Nevertheless, for $q\bar q$ states,  in first approximation, as the pion and $\eta$ meson, holographic approaches can provide the full LF wave-function due to momentum conservation. Therefore, we  obtain the GFFs from  GPDs at non-zero skewness so that, in the future, the GSW model can also be used to provide predictions for, e.g., the Sullivan process \cite{Sullivan:1971kd}, where integrals of GPDs enter the corresponding cross-section.
In this scenario, for example, one can consider a direct relation between the GFF $A(t)$ and the Light-Front  wave-function
 \cite{Brodsky:2008pf,Brodsky:2000ii,Abidin:2008hn}:

\begin{align}
    A(t) = \sum_n \int   \left[ \prod_{i=1}^n dx_i d \vec k_{\perp,i} \right] \delta \left(1-\sum_{l=1}^n x_l \right) \delta^{(2)}\left(\sum_{l=1}^n  \vec k_{\perp,l}\right)  \sum_{j=1}^n  x_j \psi^*_n(x_i, \vec k'_{\perp,i})\psi_n(x_i, \vec k_{\perp,i})~,
    \label{Eq:A_LF}
\end{align}
where $n$ is the number of parton considered for a given model of the hadron, $\vec k_{\perp,i}$ is the intrinsic transverse moment of the $i$ parton, $\psi_n$ is the LF wf which provides  the probability of describing the hadron with $n$ interacting partons,  $\vec k'_{\perp,i}  = \vec k_{\perp,i}+ (1-x_i) \vec \Delta_\perp$ for the struck constituent quark and 
$\vec k'_{\perp,i}  = \vec k_{\perp,i} -x_i \vec\Delta_\perp$ for the spectators. 
Let us remind that within the LF approach \cite{Brodsky:1997de,Diehl:2000xz} one can decompose the hadron state in terms of free constituent states:

\begin{align}
\label{Eq:Fock}
 \left|\psi_H\left(P^{+}, {\vec P}_{\perp}, S_z\right)\right\rangle  &=\sum_{n} \prod_{i=1}^n \int \frac{d x_i}{\sqrt{x_i}} \frac{d \vec {k}_{\perp i}}{2(2 \pi)^3} 16 \pi^3 \delta\left(1-\sum_{j=1}^n x_j\right) \delta^{(2)}\left(\sum_{j=1}^n \vec {k}_{\perp j}\right) \\
 \nonumber
& \times \psi_{n}\left(x_i, \vec{k}_{\perp i}\right)\left|x_i P^{+}, x_i \vec{P}_{\perp}+\vec{k}_{\perp i}\right\rangle
\end{align}
where $P^+ = P^0+P^3$ and $\vec P_\perp$ are the longitudinal and transversal  components of the hadron momentum, respectively, and $|x_i P^+, x_i \vec P_\perp+\vec k_{\perp,i}\rangle$ is the free state of the $i-$ parton with longitudinal momentum  $k_i^+=x_i P^+$ and total transverse momentum $x_i \vec P_\perp+\vec k_{\perp,i}$. This approach is particularly suitable to evaluate the GFF $A(t)$ once the LF wf is obtained from a given model.
 For example, from Eq. (\ref{Eq:poly}):

\begin{align}
    A_a(t) = \int_{-1}^1 dx~x H^a(x,0,t)~. 
\end{align}
We recall that since in the forward limit, i.e. $\xi = t=0$, the GPD $H^a(x,0,0)$ reduces to the usual PDF  $f^a(x)$, one obtains:

\begin{align}
    \sum_a A_a(0) = \sum_a \int_{-1}^1 dx~x f^a(x)=1
\end{align}
due to momentum conservation.
Unluckily,   a direct relation between the GFF $D(t)$ and  the corresponding LF wave-function is not available. Nevertheless,   one can extract $D(t)$ from
the polynomial condition Eq. \eqref{Eq:poly}:

\begin{align}
    D_a(t)= \frac{1}{2} \dfrac{\partial^2}{\partial \xi^2}\int_{-1}^1dx~ x H^a(x,\xi,t) \Big |_{\xi=0}~.
    \label{Eq:strat}
\end{align}
Hence, once the  GPD $H$ at non-zero skewness is evaluated,
the $D(t)$ GFF can be obtained.
To this aim, in the next section, we introduce the pion and $\eta$ corresponding wave-functions obtained from the GSW model.
Before closing this section, we  introduce the mean radii that can be extracted from GFFs.  The matter radius:

\begin{align}
    \langle r^2 \rangle_{mat} &= -6 \left. \frac{d A(t)}{dt} \right |_{t=0},
    \label{Eq:matt_radius}
\end{align}
and the mechanical radius defined as in Ref. \cite{Polyakov:2018zvc,Freese:2021czn}:

\begin{align}
   \langle r^2 \rangle_{mech} = \dfrac{6 D }{\displaystyle \int_{-\infty}^0 dt~D(t) } ,
   \label{Eq:mech_radius}
\end{align}
 where $D\equiv D(0)$. We also will test the conjecture that $\langle r^2 \rangle_{mech} \geq -9 D/(4 m^2)$ \cite{Polyakov:2018zvc}.

\section{Pseudo-scalar wave-functions within the GSW model}
In this section, we introduce the holographic gravitational soft-wall model \cite{Rinaldi:2017wdn,Rinaldi:2018yhf,Rinaldi:2020ssz,Rinaldi:2021dxh,Rinaldi:2022dyh,Rinaldi:2024fgx} as an improvement of the so-called AdS/QCD
 Soft-Wall  model
 \cite{Brodsky:2007hb,Karch:2006pv}.  The initial  goal of the approach was the description of the 
 scalar and tensor 
glueball spectra 
 obtained
 from the mode function of gravitons propagating
on a deformed  $AdS_5$ space \cite{Rinaldi:2017wdn}:

\begin{equation}
ds^2=\frac{R^2}{z^2} e^{\alpha k^2 z^2} (dz^2 + \eta_{\mu \nu} 
dx^\mu dx^\nu) =  e^{\alpha {k^2 z^2}} g_{M N} dx^M 
dx^N  =\bar{g}_{MN}dx^M dx^N.
\label{metric5}
\end{equation}
The parameter $k \sqrt{\alpha}=0.37 $ GeV
has been fixed to fit the slope of the glueball spectrum.
Later on, the approach was properly modified to predict the spectra of other mesons. 
To this aim we introduced the following action: \cite{Rinaldi:2021dxh}:

\begin{align}
\label{Eq:action}
 S = \int d^5x~ e^{-\varphi_0(z)} e^{-\varphi_n(z)} \sqrt{-g} \Big[g^{MN} \partial_M \Phi(x)
\partial_N \Phi(x)+M_5^2 R^2 e^{\alpha k^2 z^2} \Phi(x)^2 \Big]~,
\end{align}
where $\Phi(x)$ is the scalar field propagating in this distorted space and   $\varphi_0(z)=k^2 z^2$ is the dilaton profile function introduced in the SW model \cite{Brodsky:2007hb,Karch:2006pv}.  
For pseudo-scalar fields, 
 the 
 five dimensional 
mass is $M_5^2 R^2 = -4$~\cite{Contreras:2018hbi}. Finally,
  $\varphi_n(z)$ is an additional dilaton introduced to ensure that the effective potential entering the equation of motion is binding \cite{Rinaldi:2021dxh}. 
We emphasize that this new quantity does not take into account any further free parameter. By comparing  our calculations with the light scalar spectrum, 
the parameter  $\alpha$ in Eq. \eqref{Eq:action} has been fixed to
 $\alpha = 0.55 \pm 0.04$  \cite{Rinaldi:2020ssz}.

Thanks to  the good agreement between  the GSW predictions for scalar glueball and meson data, the model has been  tested to reproduce the spectra of other mesons without invoking  additional free parameters 
 \cite{Rinaldi:2021dxh}.
 Remarkable agreements with the current experimental state of the art have been addressed  \cite{Rinaldi:2021dxh}. 
 Building upon the predictive capabilities of the GSW model, we further explored various properties of the pion
 in Ref. \cite{Rinaldi:2022dyh} where, also in this case, data of several observable have been reproduced.

Consequently, in this investigation, we extend the study of the partonic inner structures of the pion and $\eta$ meson to the GFFs and other distributions related to
 the mechanical properties of these two pseudo-scalar mesons.
 In the next sections we introduce the $\eta$ and pion wave-functions.

\subsection{The $\eta$ wave-function}
In order to obtain the $\eta$ wf within the GDW model, one needs to solve 
the equation of motion, given the  action Eq. \eqref{Eq:action}, which can be recast in a \sch~ like equation in terms of the 5th variable $z$ \cite{Rinaldi:2021dxh}:

\begin{equation}
-\frac{d^2 \chi^\eta(z)}{dz^2} +V_\eta(z)  \chi(z) = M_\eta^2 \chi^\eta(z),
\label{psmeq0}
\end{equation}
where  $M_\eta$ is  the mass of the $\eta$ meson and the following decomposition of the scalar field is assumed:

\begin{align}
\label{Eq:TR2}
      \Phi^\eta(z)= e^{(\varphi_0(z)+\varphi_n(z))/2}~ \chi^\eta(z)z^{3/2} .
\end{align}

As shown in Refs. \cite{Rinaldi:2021dxh,Rinaldi:2022dyh}, thanks to additional dilaton, the potential for the scalar field in the above equation is:

\begin{equation}
V_\eta(z) =  k^4z^2 (1-2 \alpha^2) + 2 k^2 (1-2\alpha) - \frac{1}{4z^2} 
\label{psmeq}
\end{equation}

and the corresponding  solution is:

\begin{align}
 \chi^\eta_n(z)
&=N_n z^{ \frac{1}{2} } e^{-\frac{\sqrt{1-2 \alpha^2}}{2}k^2 z^2  } ~ _1F_1 \left(-n,1, 
\sqrt{1-2 \alpha^2} k^2 z^2
 \right)~.
 \label{etawf}
\end{align}
 $N_n$ is a normalization constant, $_1 F_1$ the hypergeometric confluent function and the $\eta$ mass spectrum is given by:

\begin{align}
 M_\eta(n)^2 =  \Big[(2-4 \alpha)+2\sqrt{1-2\alpha^2}(1+2 n)\Big]k^2.
\end{align}
One should  notice that
 $M_\eta(0) \ne 0$, i.e., the model itself does not predict
a massless ground state.
Since we obtained $M_\eta  = 513 \pm 92$  MeV \cite{Rinaldi:2021dxh}. 
In fact, our prediction is compatible with the experimental value: $M_\eta \sim 547.862 \pm 0.017$ MeV. We remind that the theoretical uncertainty in the GSW model is due to  the fit of $\alpha$  \cite{Rinaldi:2021dxh}. 
The wave-function can be obtained through the procedure described, in e.g., Refs. \cite{Brodsky:2007hb,Rinaldi:2022dyh}, where the density distribution has been introduced:

\begin{align}
 \tilde \rho_\eta(x,z) = \dfrac{e^{-\varphi_0(z) -\varphi_n(z)} }{2 \pi} \dfrac{x}{1-x} 
\dfrac{ |\Phi^\eta(z)|^2 }{z^4} =  \dfrac{x}{1-x} 
\dfrac{ |\chi^\eta(z)|^2 }{2 \pi z}  =
 \dfrac{|\tilde \psi_{\eta}(x, \vec b_\perp) |^2}{(1-x)^2}~,
\end{align}
here 
${\vec b}_\perp$ represents the transverse distance between the quark and the anti-quark, conjugate variable to $\vec k_\perp$ and $\tilde \psi_{\eta}$ is the FT of $\psi_2$ in Eq. \eqref{Eq:Fock} (for the $q \bar q$ state).
Therefore, we get the following correspondence:

\begin{align}
 \tilde \psi_{\eta}(x, \vec{b}_\perp)
=
\sqrt{\dfrac{x(1-x) }{2 \pi z}  } \chi^\eta_0(z)~,
\label{AdsLF}
\end{align}
Finally, the AdS/QCD approach requires the following identification:
 $z =\sqrt{x(1-x)} |\vec{b}_\perp|$ \cite{Brodsky:2007hb}. 
 In order to  test the model, we display 
 in the left panel of Fig. \ref{fig:TFF}, the $\eta$ transition form factor (TFF), compared with data \cite{BaBar:2011nrp,CELLO:1990klc}.  As one can see, data are well reproduced  \cite{Ahmady:2018muv} without adding further free parameters. We also display, in the right panel of the same figure, the pion TFF.
 The latter can be obtained from the pion wf \cite{Rinaldi:2022dyh} discussed in details in the next section.
As one can see, the GSW predictions compare well with the TFF data for both these two pseudo-scalars.

\subsection{The pion wave-function}
As discussed in Ref. \cite{Rinaldi:2022dyh}, to attempt embedding the chiral symmetry-breaking mechanism within the GSW model, we proposed a two-step procedure. First, as in the soft-wall model of Ref. \cite{Brodsky:2007hb}, we required  the model fulfilment of the chiral symmetry, meaning a massless  pion ground state. 
To this aim, we imposed that the additional dilaton in Eq. \eqref{Eq:TR2} leads to the following potential in the \sch~  
Eq. \eqref{psmeq0}:

\begin{align}
V_\pi(z) = \frac{15}{4 z^2} +2 k^2+k^4 z^2 - \frac{4}{z^2}\left[1+(\alpha 
+ \xi_\pi)k^2 z^2+\frac{1}{2}( \alpha^2 + \gamma_\pi) k^4 z^4 \right] 
= V_\eta(z)-4 k^2 \xi_\pi-2 \gamma_\pi k^4 z^2~.
\label{pionpotential}
\end{align}
where the parameter 
 $\gamma_\pi$ modifies the $\eta$ slope of the spectrum to reproduce that of  the pion and $\xi_\pi$ leads to the massless ground state:

\begin{align}
 M^2_\pi(n) = 4\sqrt{1-2 (\alpha^2+ \gamma_\pi)}~k^2n ~.
\label{mpi} 
\end{align}
see Ref. \cite{Rinaldi:2022dyh} for details. Similarly to Eq. \eqref{etawf}, the corresponding mode function is:

\begin{align}
 \chi^\pi_n(z)
&=N_n z^{ \frac{1}{2} }~ exp\left(-\frac{\sqrt{(1-2 (\alpha^2- \gamma_\pi)}}{2}k^2 z^2  \right) ~ _1F_1 \left(-n,1, 
\sqrt{(1-2 (\alpha^2+\gamma_\pi)} k^2 z^2
 \right)~.
 \label{pionwf}
\end{align}
As a second step, we followed the line of Refs. \cite{tHooft:1974pnl,Li:2015zda,Burkardt:1997de,Li:2021jqb,deTeramond:2021yyi}, where an additional longitudinal dynamics is taken into account to effectively describe the chiral symmetry
breaking mechanism \cite{tHooft:1974pnl}.
In this framework, one introduces an effective potential  
$V_{eff} = V_\perp+V_{||}$,  where $V_\perp = V_{\pi}$ in Eq. \eqref{psmeq0} and 
$M_\pi^2 = M^2_\perp +M_{||}^2$, given $M^2_\perp=M_\pi^2$ in Eq. \eqref{mpi}. In order to get the longitudinal part, the following mass equation should be solved:

\begin{align}
\left[  \dfrac{m_q^2}{x} + \dfrac{m_{\bar q}^2}{1-x}+V_{||}(x) \right] 
\Xi(x)&=M^2_{||}~ \Xi(x)~,
\label{Eq:long}
\end{align}
where he $\Xi(x)$ is the longitudinal component of the  total wave-function, which reads:

\begin{align}
\tilde \psi_\pi(x, \vec{b}_\perp) = 
\tilde\psi^\perp_{\pi}(x,\vec{b}_\perp)~ \Xi(x)~,
\label{LFwf}
\end{align}
 where, similarly for the $\eta$ wf, the "transverse" 
 wave-function is:
 
 \begin{align}
     \tilde \psi^\perp_\pi(x,\vec b_\perp) =  \sqrt{\frac{x(1-x)}{2 \pi z}} \chi^\pi_0(z)
 \end{align}
 
\begin{figure*}
\centering
\includegraphics[scale=0.5]{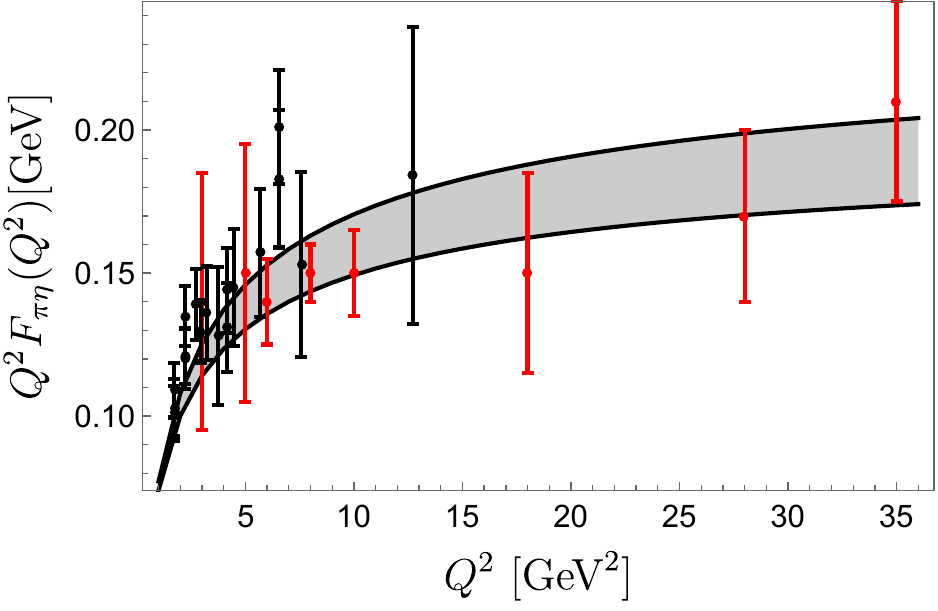}
\includegraphics[scale=0.52]{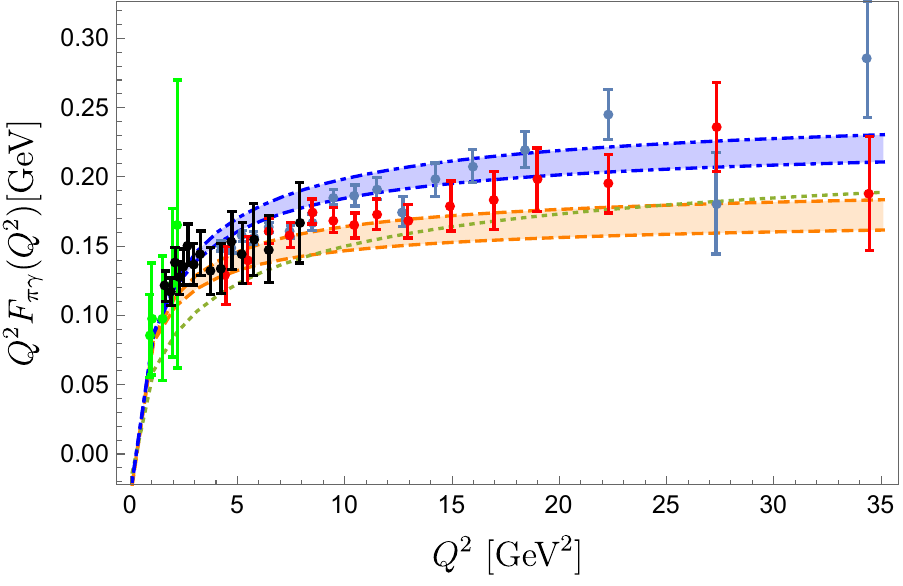}
\caption{\footnotesize  {
Left panel:
the $\eta$ TFF compared with data \cite{BaBar:2011nrp,CELLO:1990klc}. Right panel: 
the pion TFF from Ref. \cite{Rinaldi:2022dyh}. Dashed lines for the GSWL2 set and dot-dashed lines for GSWL1 one. Data are from Refs. \cite{CELLO:1990klc,Belle:2012wwz,BaBar:2009rrj,CLEO:1997fho}. 
Bands
stand for the error in $\alpha$. 
}}
\label{fig:TFF}
\end{figure*}

\begin{figure*}
\centering
\includegraphics[scale=0.59]{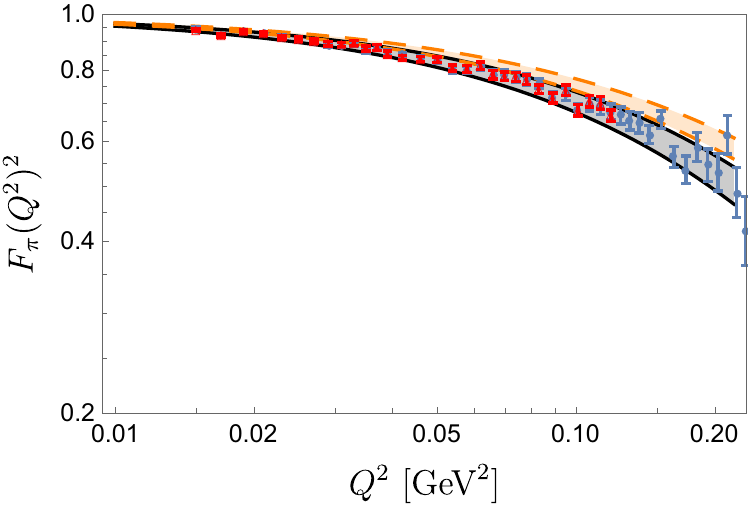}
\caption{\footnotesize  {The pion FF from Ref. \cite{Rinaldi:2022dyh}. Full lines for the GSWL2 model and dashed 
lines for the GSWL1 model. Data from Refs. 
\cite{NA7:1986vav,Ackermann:1977rp,Bebek:1977pe,JeffersonLabFpi:2007vir,Brauel:1979zk,JeffersonLabFpi-2:2006ysh,Amendolia:1984nz}.
Bands
stand for the error in $\alpha$.  
}}
\label{fig:ff}
\end{figure*}

  In Ref. \cite{Rinaldi:2022dyh}, we used the longitudinal potential model of Refs. 
\cite{Li:2021jqb,Li:2015zda,Li:2022izo}:

\begin{align}
 V_{||}(x)=-\sigma^2 \partial_x \big[x(1-x)\partial_x \big]~,
 \label{longpot}
\end{align}
 where $\sigma$ parametrizes the strength of 
confinement~\cite{Li:2021jqb} and can be fixed to reproduce the experimental value of the pion ground state mass.
The total solution  is:

\begin{align}
 M^2_\pi = M^2_{\perp \pi}+ \sigma (m_q+m_{\bar q} )(2l+1)+(m_q+m_{\bar q} )^2+ 
\sigma^2l(l+1)~,    
\label{mpilong}
\end{align}
with  the longitudinal wave-function:
\begin{align}
\Xi(x)=N_2 x^{\beta/2} (1-x)^{\alpha/2}P_l^{(\alpha,\beta)}(2x-1)~,
\label{Eq:chi}
\end{align}
$\alpha = 2 m_q/\sigma$, $\beta = 2 m_{\bar q}/\sigma$~
\cite{Li:2021jqb} and

\begin{align}
 \sigma = \frac{M^2_{exp}-4 m_q^2}{2 m_q}~.
\label{Eq:sigma}
\end{align}
We remind that in this framework, 
 the  constituent quark masses
 must be properly fixed
\cite{Brodsky:2007hb,Li:2021jqb}.
For the moment being,
we assume $m_q = m_{\bar q}$ for light quarks. In conclusion, in order to properly include the underlying  pion dynamics, both $\gamma_\pi$ and $m_q$ must be fitted.  
We recall some results presented in Ref. \cite{Rinaldi:2022dyh} given two parametrizations of $\gamma_\pi$ and $m_q$.
We called GSLW1 (GSLW2) the set $m_q = 45$MeV ($m_q=52$ MeV) and $\gamma_\pi=-0.6$ ($\gamma_\pi= -0.17$). For example, in Tab. \ref{piontab} we report the pion spectrum compared with the PDG \cite{ParticleDataGroup:2018ovx,ParticleDataGroup:2022pth}. As one can see both parametrizations agree quite well with the present experimental scenario. We also show in Fig. \ref{fig:ff}  the low $Q^2$ behaviour of the form factor. As one can see, also in this case, both the sets of parameters reproduce the data. We also mention that the mean pion charge radius is: $\sqrt{\langle r^2 \rangle_{GSLW1}} = 0.67 \pm 0.03$ fm ($\sqrt{\langle r^2 \rangle_{GSLW2}} = 0.70 \pm 0.05$ fm) to be compared with $\sqrt{\langle r^2 \rangle_{exp}} = 0.67 \pm 0.01$ fm. Since, one of the main observable we aim to reproduce for the $\eta$ is the transition form factor, in order to compare pion and $\eta$ calculations of this quantity, we report in the right panel of Fig. \ref{fig:TFF}   the pion TFF. In all these cases, the present approach is able to describe the data for all these observable, confirming the good predicting power of the model.

\begin{table}[htb]
\centering
\begin{tabular}{| c | c | c | c | c | c | c|}
\hline
& $\pi^0$ & & $\pi(1300)$ & & & $\pi(1800)$    \\ \hline
PDG & $ 134.9768\pm 0.0005$ &  & $1300\pm100$ &  & & $1819\pm 10$  \\ \hline
SW \cite{Brodsky:2007hb} &   $ 0  $    & &       $ 1080 $& $  1527
$ &  & $1870$\\ \hline
Ref. \cite{Rinaldi:2021dxh} &   $ 135  $    &$943 \pm 111$&       $ 1231  \pm 133$& $  1463 
\pm151$ & $1663 \pm168$ & $1842\pm183$\\ \hline
GSWL1 &   $ 140  $  &      & $1199\pm41 $&   &   & $1800 \pm 6$\\ 
\hline
GSWL2 &   $ 140  $  &      & $1019 \pm 27 $&   &   & $1793 \pm 16 $\\ 
\hline
Ref. \cite{Li:2021jqb} & $140$  & &1520 & & &2120 \\
\hline
\end{tabular}
\caption{ \footnotesize The $\pi$ mass spectrum in MeV. Results from  the PDG 
particle listings~\cite{Tanabashi:2018oca,Zyla:2020zbs} compared with  those obtained from 
the   GSWL1 and GSWL2  parametrizations. The error source in the GSWL scenarios is
associated to $\alpha = 0.55 \pm 0.04$.  .}
\label{piontab}
\end{table}

\section{Meson GPDs}
In this section, we present the procedure to evaluate the pseudo-scalar GPDs from the light-cone wave-functions given by the GSW model. Let us remark that, within these holographic approaches,  the wf depends on $x$ and $k_\perp$ of one parton and GPDs at zero skewness can be evaluated following, e.g., the line of Refs. \cite{deTeramond:2018ecg,Vega:2010ns}, 
while  GPDs at $\xi \neq 0$  within the SW model can be found in Ref. \cite{Kaur:2018ewq}.
For the proton case, there are some attempts to include a dependence on the all momenta of partons
 but these procedures invoke extra modelizations, see, e.g., Ref. \cite{Rinaldi:2017roc}. In this analysis, we consider a suitable procedure to evaluate the GPDs at $\xi \neq 0$ of mesons,  described by the first $q \bar q$ valance Fock state in Eq. \eqref{Eq:Fock}. In this case, one might notice that, due to momentum conservation  one has that $x_2 = 1-x_1 = 1-x$ and $\vec k_{\perp,2} = - \vec k_{\perp,1} = -\vec k_\perp$. Therefore, since both the pion and $\eta$ wave-functions are both symmetric in $x$ and $1-x$, see Eqs. \eqref{AdsLF}-\eqref{Eq:chi} (for $l=0$ grounstate), one obtains the full wave-function:

\begin{align}
   \psi_\tau(x, \vec k_\perp)=  \psi_\tau(x_1,x_2,\vec k_{\perp,1},\vec k_{\perp,2}) = 
    \psi_\tau(x,1-x,\vec k_{\perp},-\vec k_{\perp}) 
\end{align}
where $\tau=\pi,\eta$ and $\psi_\tau(x,k_\perp)$ is the Fourier Transform of the wf of Refs. \eqref{AdsLF}-\eqref{LFwf}, respectively. 

Thanks to  this feature, the total wf of the meson $\tau=\pi,\eta$ is obtained and then the  GPD  $H_\tau(x,\xi,t)$ can be expressed as follows \cite{Boffi:2002yy}:

\begin{align}
    H_\tau(x,\xi,t) = \int \frac{d \vec k_\perp}{16 \pi^3} \psi_\tau(y'_1,y'_2,\vec \kappa'_{\perp,1}, \vec \kappa'_{\perp,2})
    \psi_\tau(y_1,y_2,\vec \kappa_{\perp,1}, \vec \kappa'_{\perp,2}),
    \label{Eq:gpdH}
\end{align}
where for the final struck quark:

\begin{align}
   \left \{  \begin{aligned}
   &y_1' = \frac{x_1-\xi}{1-\xi} 
   \\
   ~
   \\
  & \vec \kappa'_{\perp,1} = \vec k_\perp + \frac{1}{2} \frac{1-x}{1-\xi} \vec \Delta_\perp
   \end{aligned}
   \right.,
\end{align}
for the final spectator:

\begin{align}
   \left \{  \begin{aligned}
   &y_2' = \frac{1-x}{1-\xi} 
   \\
   ~
   \\
  & \vec \kappa'_{\perp,2} = -\vec k_\perp - \frac{1}{2} \frac{1-x}{1-\xi} \vec \Delta_\perp
   \end{aligned}
   \right.
\end{align}
For the initial struck quark:

\begin{align}
   \left \{  \begin{aligned}
   &y_1 = \frac{x_1+\xi}{1+\xi} 
   \\
   ~
   \\
  & \vec \kappa_{\perp,1} = \vec k_\perp - \frac{1}{2} \frac{1-x}{1+\xi} \vec \Delta_\perp
   \end{aligned}
   \right.
\end{align}
and, finally, for the initial spectator
\begin{align}
   \left \{  \begin{aligned}
   &y_2 = \frac{1-x}{1+\xi} 
   \\
   ~
   \\
  & \vec \kappa_{\perp,2} = -\vec k_\perp +\frac{1}{2} \frac{1-x}{1+\xi} \vec \Delta_\perp
   \end{aligned}
   \right.
\end{align}
 Let us remark that this is a crucial step for future comparison with experimental extractions of the pion GPDs.

\begin{figure*}
\includegraphics[scale=0.5]{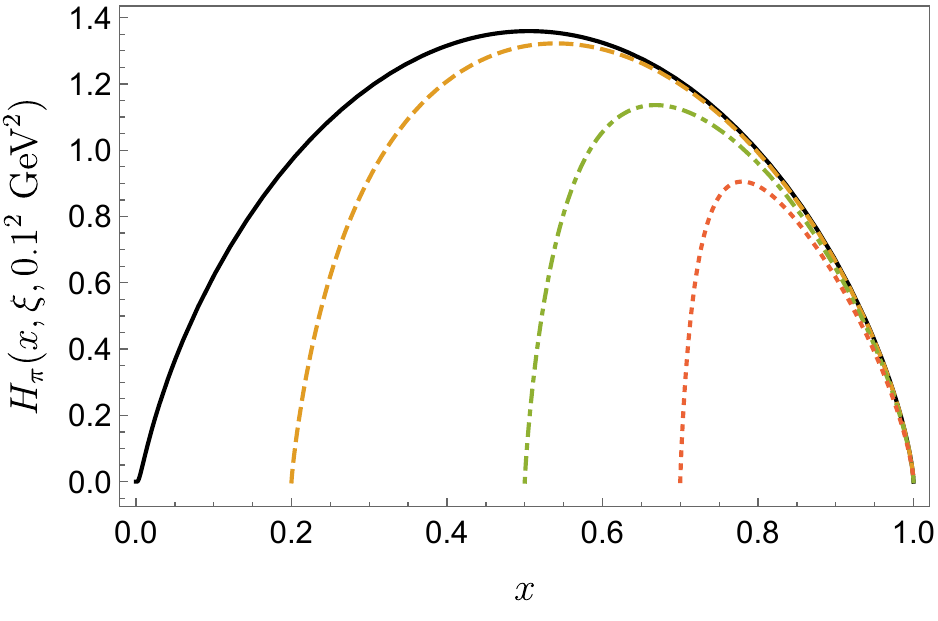}
\hskip 0.5cm 
\includegraphics[scale=0.5]{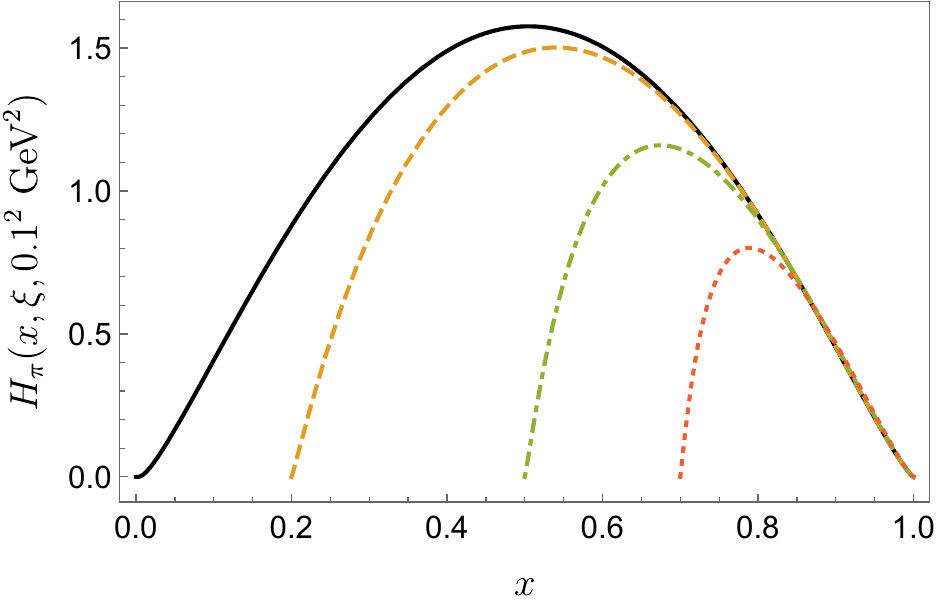}
\caption{\footnotesize  {The pion GPD $H_\pi$ defined in Eq. \eqref{Eq:gpdH} and evaluated for $t=-0.1$ GeV$^2$ as a function of $x$ and for different values of $\xi$. Full lines for $\xi=0$, dashed lines for $\xi=0.2$, dot-dashed lines for $\xi=0.5$ and dotted-lines for $\xi=0.7$. Left panel for the GSWL1 parametrization. Right panel for the GSWL2 parametrization.
}}
\label{fig:gpd_01}
\end{figure*}

\begin{figure*}
\includegraphics[scale=0.5]{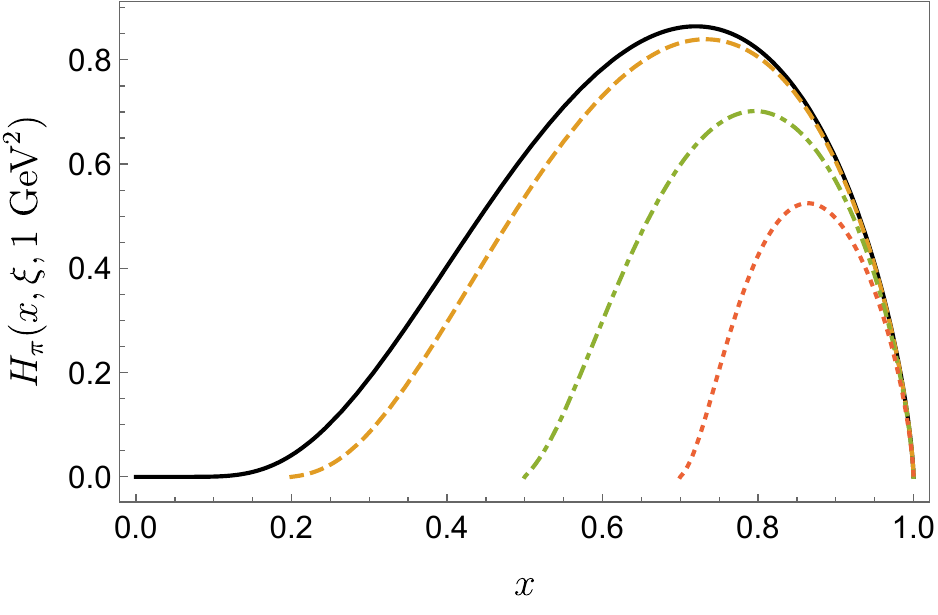}
\hskip 0.5cm 
\includegraphics[scale=0.5]{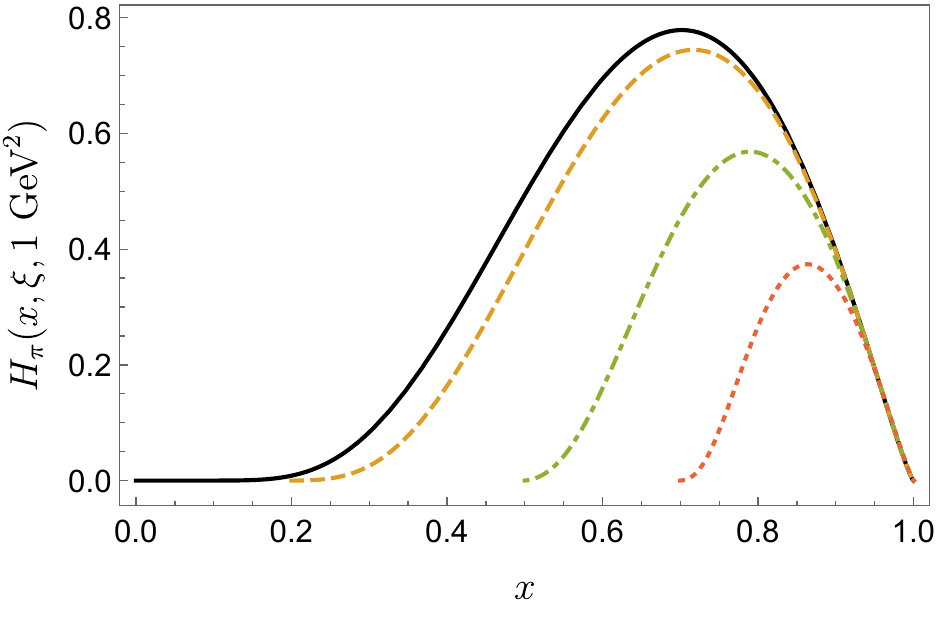}
\caption{\footnotesize  {Same of Fig. \ref{fig:gpd_01} but for $t=-1$ GeV$^2$.
}}
\label{fig:GPD_1}
\end{figure*}

\section{Results}

\begin{figure*}
\includegraphics[scale=0.5]{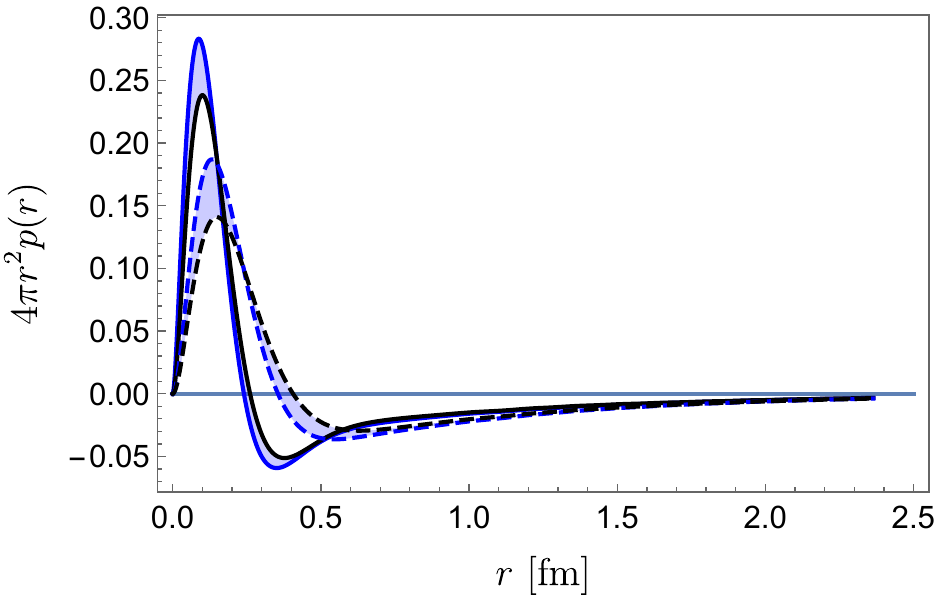}
\hskip 0.5cm 
\includegraphics[scale=0.5]{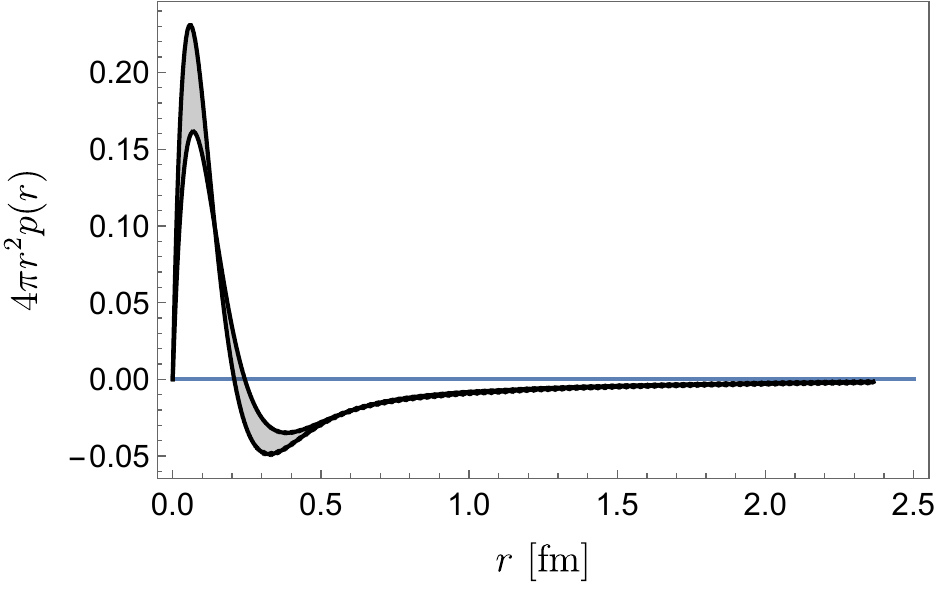}
\caption{\footnotesize  {
Left panel: the pion  pressure distribution $4 \pi r^2 p(r)$ [GeV/fm] from Eq. \eqref{Eq:p}.  Full-lines for the GSWL1 and dashed lines for the GSWL2 parametrizations, respectively.  Right panel: same of the left panel for the $\eta$. Bands
stand for the error in $\alpha$.  
}}
\label{fig:p}
\end{figure*}

\begin{figure*}
\includegraphics[scale=0.5]{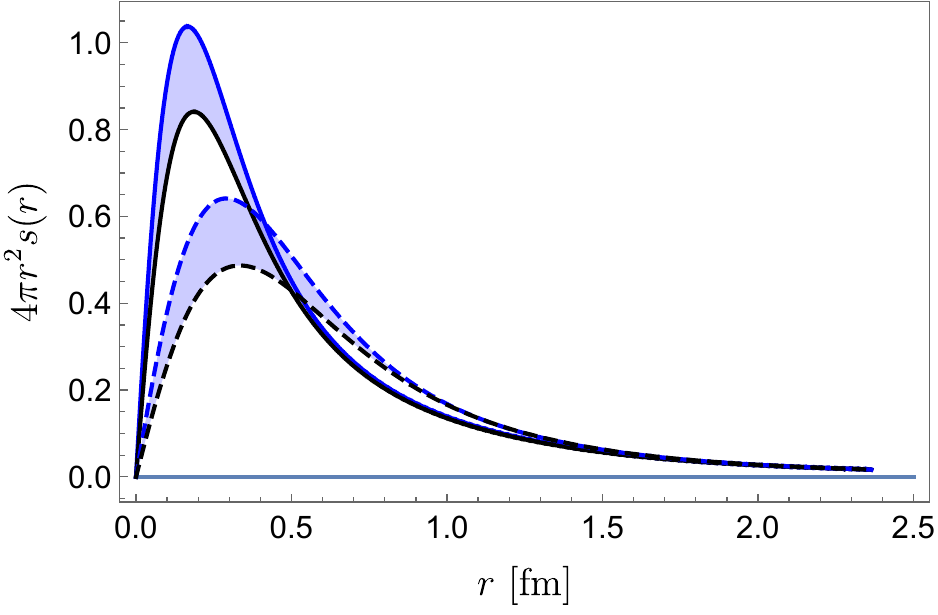}
\hskip 0.5cm 
\includegraphics[scale=0.5]{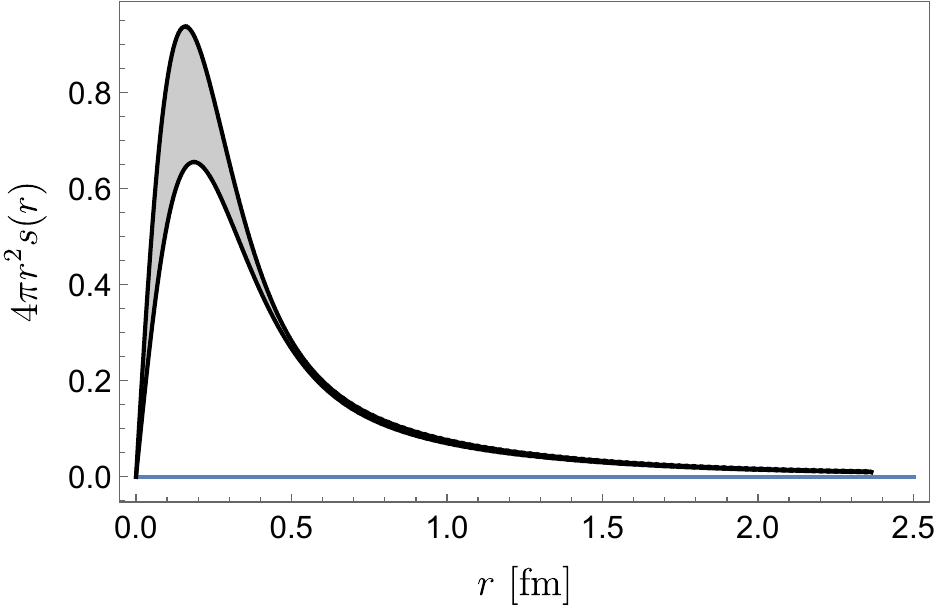}
\caption{\footnotesize  {Left panel: the pion shear-force distribution $4 \pi r^2 s(r)$ [GeV/fm] from Eq. \eqref{Eq:s}.  Full lines for the GSWL1 and dashed lines for the GSWL2 parametrizations, respectively.  Right panel: same of the left panel for the $\eta$. Bands
stand for the error in $\alpha$.  
}}
\label{fig:s}
\end{figure*}

\begin{figure*}
\includegraphics[scale=0.5]{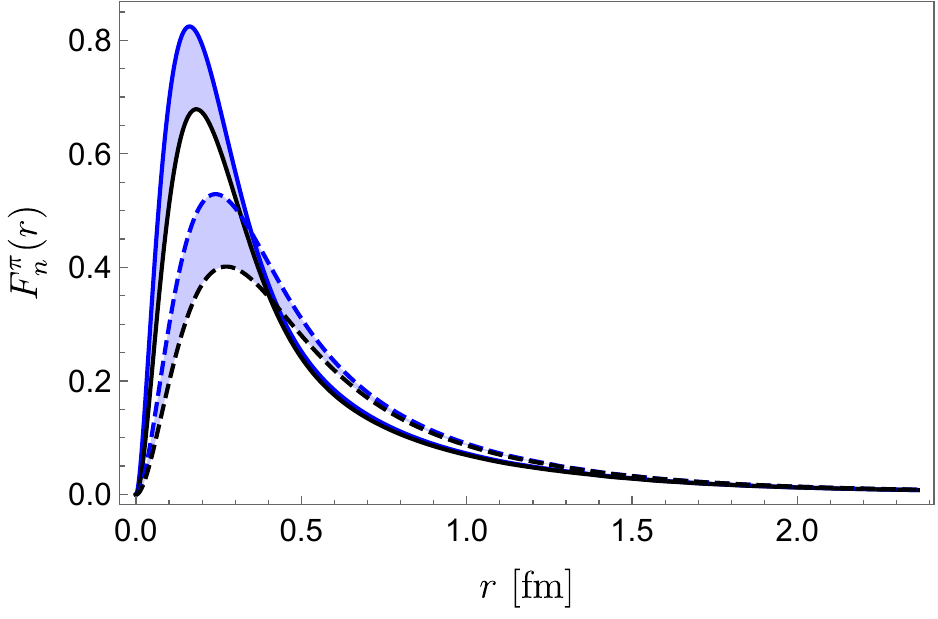}
\hskip 0.5cm 
\includegraphics[scale=0.5]{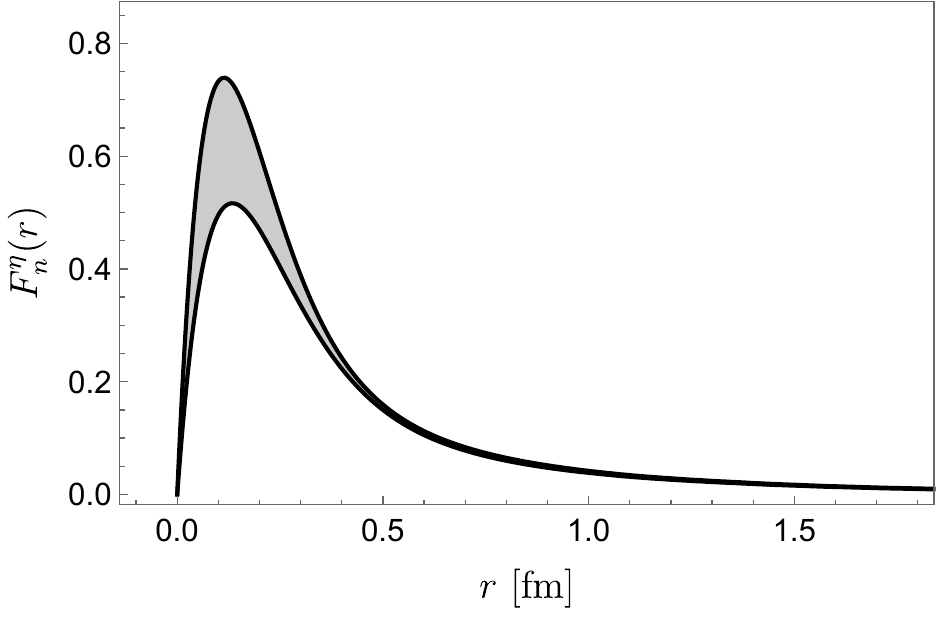}
\caption{\footnotesize  {
The normal force Eq. \eqref{Eq:Fn}. Left panel for the pion. Full lines  for the GSWL1 parametrization and the dashed lines  for the GSWL2 one. Right panel for the $\eta$.
 Bands
stand for the error in $\alpha$.  
}}
\label{fig:Fn}
\end{figure*}

\begin{figure*}
\includegraphics[scale=0.5]{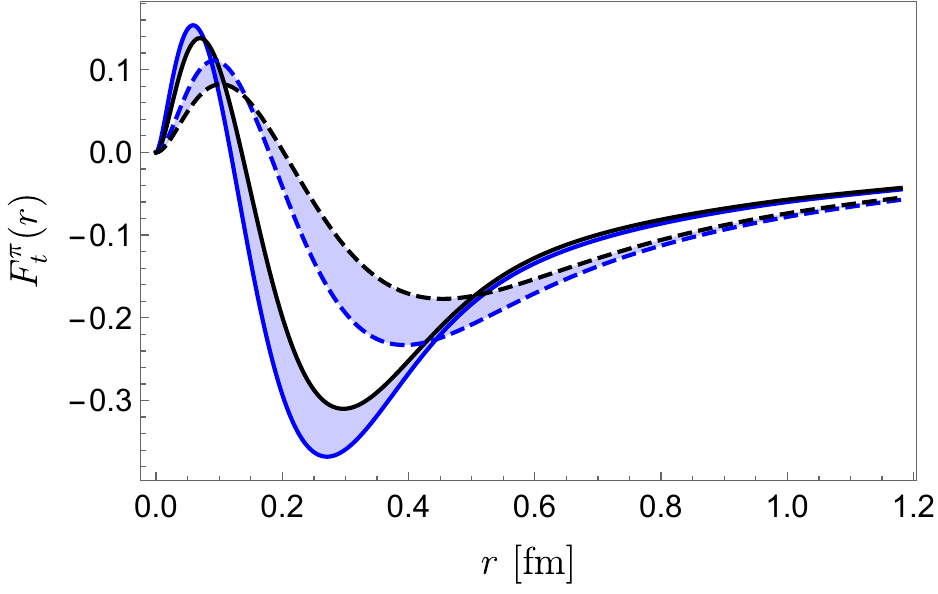}
\hskip 0.5cm 
\includegraphics[scale=0.5]{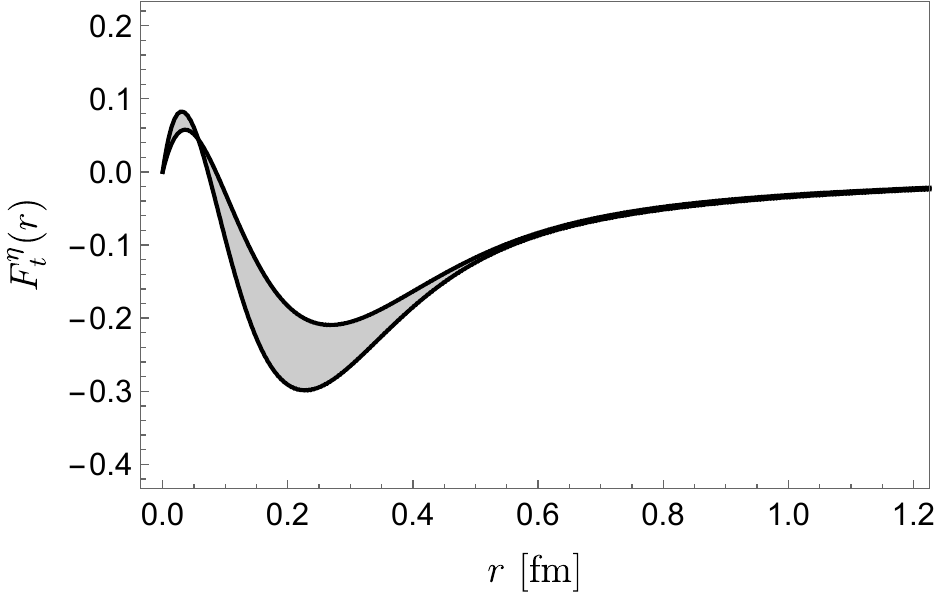}
\caption{\footnotesize  {Same of Fig. \ref{fig:Fn} for the tangential forces Eq. \eqref{Eq:Ft}.
}}
\label{fig:Ft}
\end{figure*}

 With the confidence gained from the agreement of the previous calculations with  data, we now present in Fig. \ref{fig:gpd_01}  the $x$ dependence of the pion GPD $H_\pi(x,\xi,t)$, Eq. \eqref{Eq:gpdH}, for $t=-0.1$ GeV$^2$, as a function of $\xi=0.1,~0.2,~0.5,~0.7$.
 Left (right) panel corresponds to the GSWL1 (GSWL2) parametrization.
In the  Fig. \ref{fig:GPD_1}, we report the same GPDs but evaluated at $t=-1$ GeV$^2$. 
As expected, the decreasing $t$ behaviour follows the form factor (ff) one. Moreover, the $\xi$ dependence is qualitatively in agreement with that of other analyses, for example in Ref. \cite{Chavez:2021llq}. 
In the left panel of Fig. \ref{fig:A}, we show the calculations of the pion  $A(t)$ GFF obtained from Eq. (\ref{Eq:A_LF}) and compared  with the lattice data of Ref. \cite{Hackett:2023nkr}, along the line of Ref. \cite{Li:2023izn}. As one can see, the GSWL1 set is capable of reproducing the lattice data. We remark that in Ref. \cite{Li:2023izn}, the same quantity has been evaluated within the  soft-wall model where dressed quarks have been taken into account.
In the right panel of Fig. Fig. \ref{fig:A} the pion GFF $A_\pi(t)$, obtained from the GSLW1 parametrization (full lines, is compared with the prediction of $A(t)$ for the $\eta$ meson (dashed lines).  
 As evident, the  outcomes are  different, thereby elucidating the GFF $A^\tau(t)$ sensitivity to the diverse internal structures of the two pseudo-scalar systems.
Within the same strategy, we display in Fig. \ref{fig:D} the  pion (left panel) and $\eta$ (right panel) GFFs $D^\tau(t)$.
This quantity has been obtained from the meson GPD $H_\tau$, thanks to the polynomial constraint Eq. \eqref{Eq:poly} and Eq. \eqref{Eq:strat}.
Also in this case, the model is almost capable to reproduce the lattice data \cite{Hackett:2023nkr} without any additional fit of the parameters. Moreover, the lattice predictions in the low $t$ region seem to foster  the GSWL1 parametrization. In the right panel we compare the calculation of this quantity for the pion ($\tau=\pi$) with that for the $\eta$ meson ($\tau=\eta$).
As one can see, the GSW model predicts that in the low $t$ region the $D(t)$ GFF for the two mesons are quite different, while  distributions are comparable at hight $t$.
 From  these calculations of both the GFFs $A(t)$ and $D(t)$, the matter and mechanical radii can be numerically evaluated from Eqs. \eqref{Eq:matt_radius}-\eqref{Eq:mech_radius}. We report results in Tab.  \ref{Tab:radii}. 
As one can see, both the matter and mechanical pion radii are smaller than the charge radius predicted by the model. Furthermore,  all parametrizations predict that the pion radii are also smaller then those of the $\eta$ meson. 
Results can be compared with those addressed in, e.g., Ref. \cite{Freese:2021czn}.
As final part of this section, the GFF  $D^\tau(t)$ is used to evaluate 
  both the pressure and shear-force distributions of the corresponding meson, following Eqs. (\ref{Eq:p})-(\ref{Eq:s}). 
  We display in Fig. \ref{fig:p} the pressure distributions for the pion (left panel) and the $\eta$  meson   (right panel).  By using the convention of Ref. \cite{Polyakov:2018zvc}, we can conclude that the repulsion contribution of the interaction, i.e.  where the distribution is positive, is higher in the pion then that in the $\eta$. This is a pure prediction of the GSW model. 
  We also report in Fig. \ref{fig:s} the shear-force distributions for the pion (left panel) and the $\eta$ (right panel). 
  We recall that this quantity must be non-vanishing to ensure a finite mechanical radius and, consequently, the formation of inner hadronic structure.
Furthermore,  the shear-force distribution is more sensitive to GSW parameters w.r.t. the pressure distribution. 
We conclude this first section of results
displaying and discussing 
 the normal and tangential force distributions felt by a spherical shell of radius $r$. These quantities, for $\tau=\pi$ and $\tau=\eta$, can be defined as two different combinations of both pressure and shear-force distributions \cite{Polyakov:2018zvc}:

\begin{align}
\label{Eq:Fn}
    F^\tau_n(r) &= 4 \pi r^2 \left[\frac{2}{3} s^\tau(r)+p^\tau(r) \right]
    \\
    F^\tau_t(r) &= 4 \pi r^2 \left[-\frac{1}{3} s^\tau(r)+p^\tau(r) \right].
    \label{Eq:Ft}
\end{align}
We recall that within the present convention, positive forces correspond to \textit{stretching}  effects while negative values are related to \textit{squeezing} effects. In Fig. \ref{fig:Fn} we report the normal force distribution, Eq. \eqref{Eq:Fn}, evaluated for the pion (left-panel) and for the $\eta$ meson (right-panel). One should notice that, as expected, the stability condition leading to $F_n(r)>0$ \cite{Polyakov:2018zvc,Perevalova:2016dln} is fulfilled. 
Moreover, comparing our results with those of Ref. \cite{Polyakov:2018zvc}, we conclude that the present model predicts a normal force intensity, for these mesons,  higher than that of the nucleon predicted by the quark-soliton model \cite{Goeke:2007fp}.
In Fig. \ref{fig:Ft} we display the tangential force distribution (cf. Eq. \eqref{Eq:Ft}) for the pion (left-panel) and for the $\eta$ meson (right-panel). In this case the stretching contribution is similar for both mesons and similar to that predicted by the quark-soliton model for the nucleon; however the squeezing contribution is more then 3 times higher then that for the nucleon \cite{Goeke:2007fp}.

\section{GFF from DA and comparison with lattice calculations}
\begin{figure*}
\includegraphics[scale=0.34]{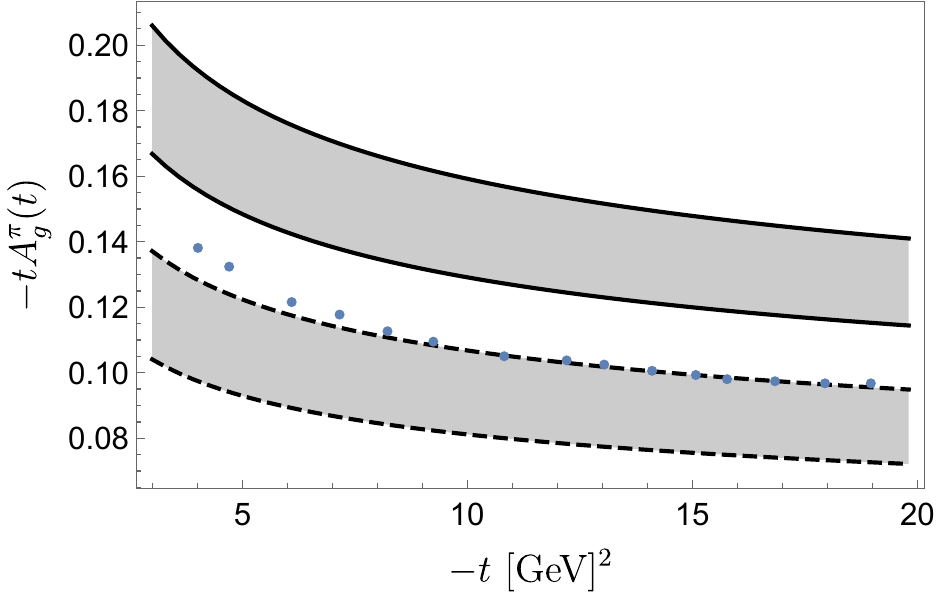}
\includegraphics[scale=0.34]{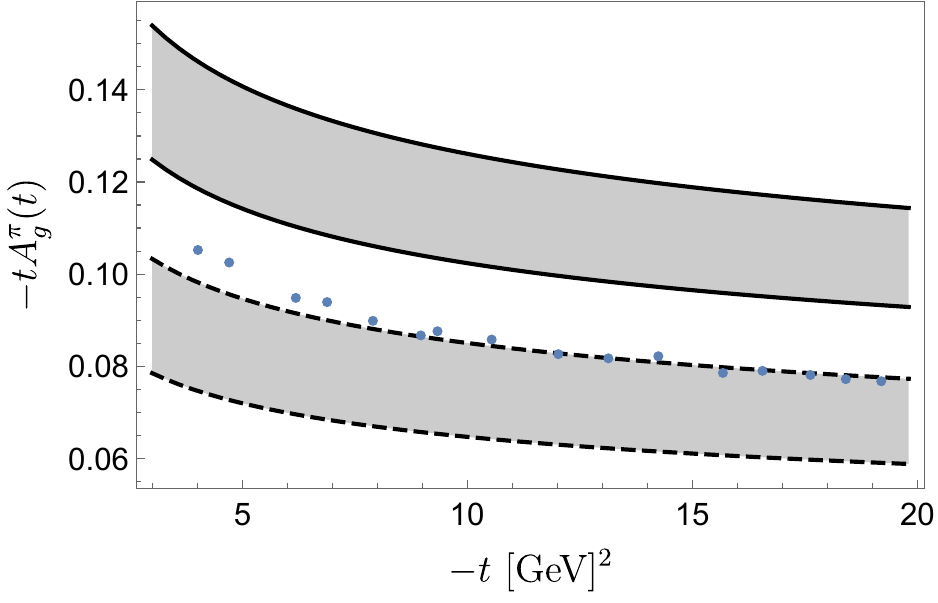}
\includegraphics[scale=0.34]{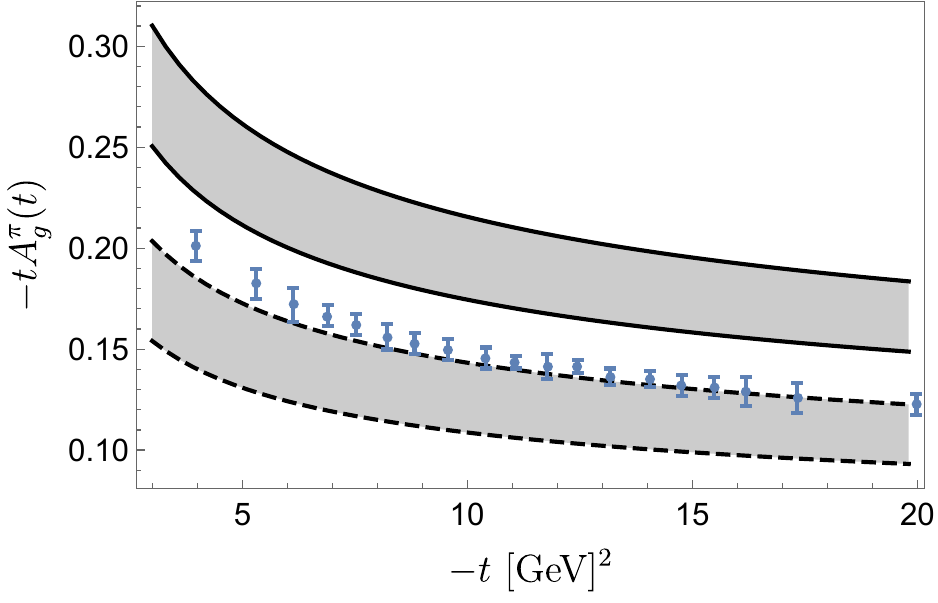}
\caption{\footnotesize  {The pion GFF $q^2 A^\pi_g(q^2)$ evaluated within Eq. 
\eqref{Eq:Agl}
\cite{Gao:2022vyh}. Full lines for the GSWL1 parametrization and dashed lines for the GSWL2 one.
Bands
stand for the error in $\alpha$.  
Left panel for $\mu_R^2 = q^2$. Central panel for $\mu_R^2=4 q^2$. Right panel for $\mu_R^2 = q^2/4$. Momentum transfert in GeV$^2$.
}}
\label{fig:DA_Ag}
\end{figure*}

\begin{figure*}
\includegraphics[scale=0.34]{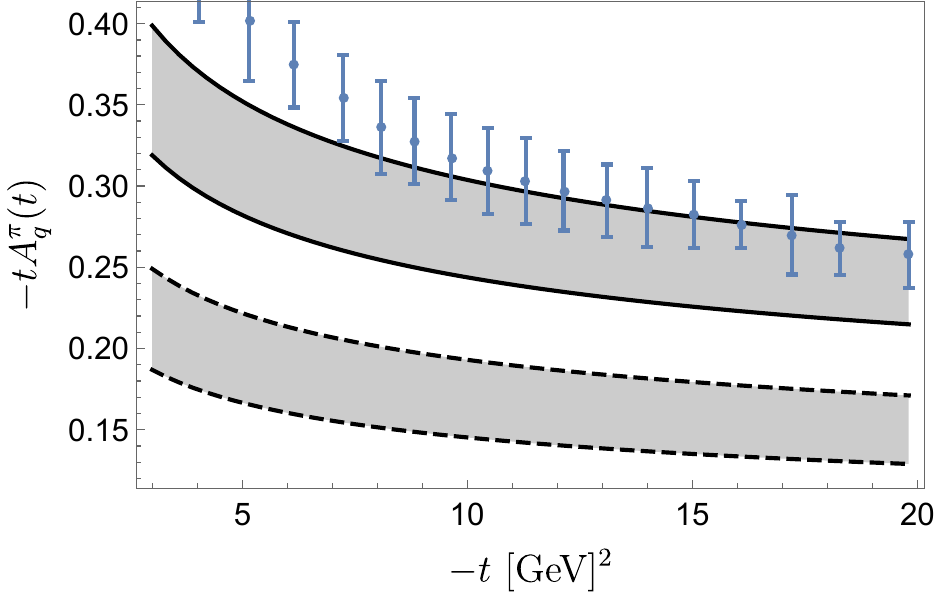}
\includegraphics[scale=0.34]{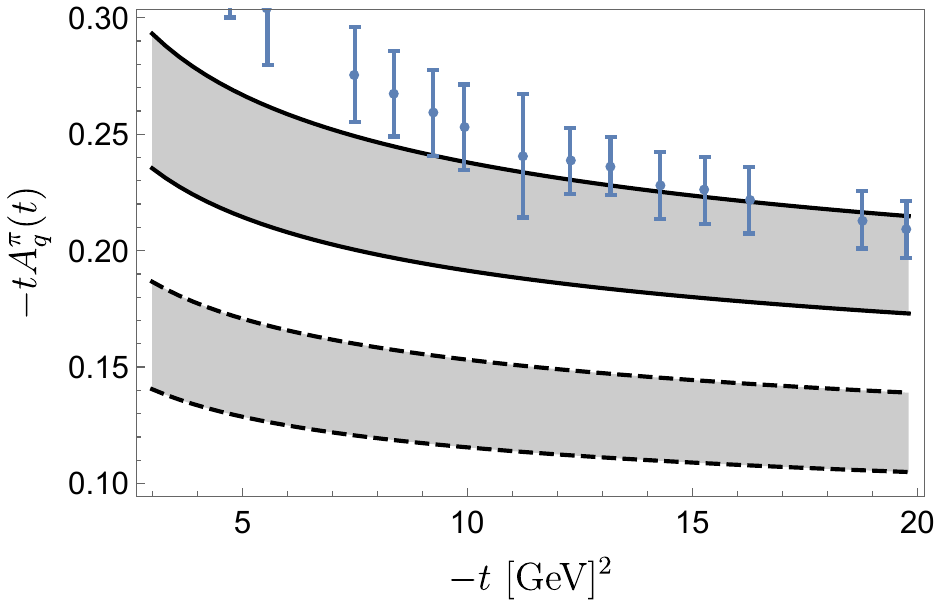}
\includegraphics[scale=0.34]{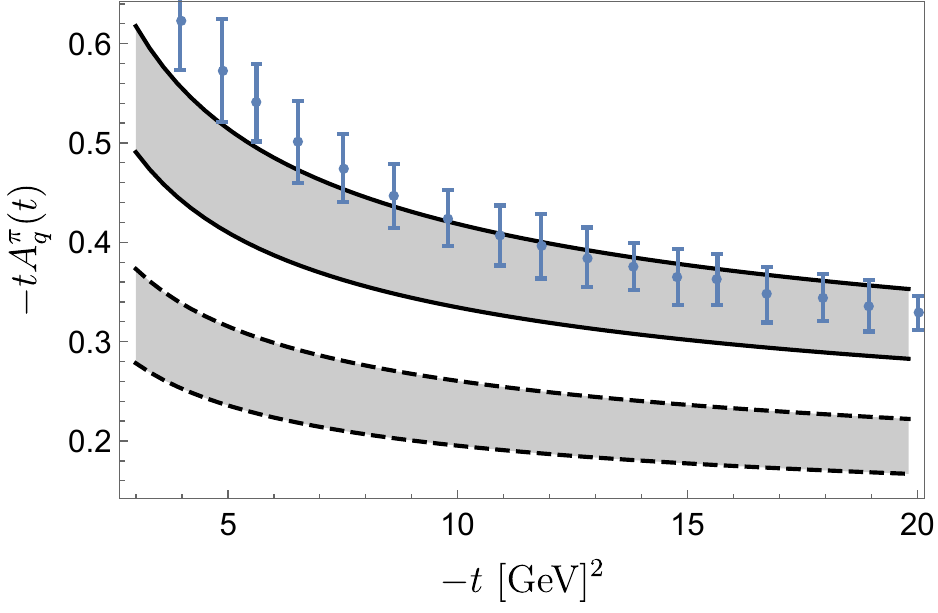}
\caption{\footnotesize{
Same of Fig. \ref{fig:DA_Ag} for
the pion GFF $q^2 A^\pi_q(q^2)$ evaluated within Eq. \eqref{Eq:Aql} \cite{Gao:2022vyh}. 
}}
\label{fig:DA_Aq}
\end{figure*}

\begin{figure*}
\includegraphics[scale=0.34]{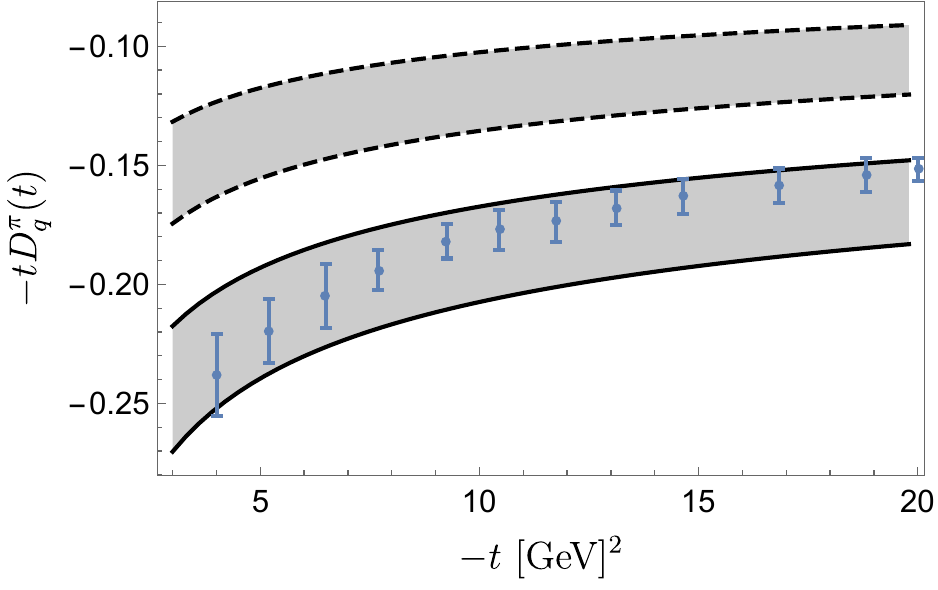}
\includegraphics[scale=0.34]{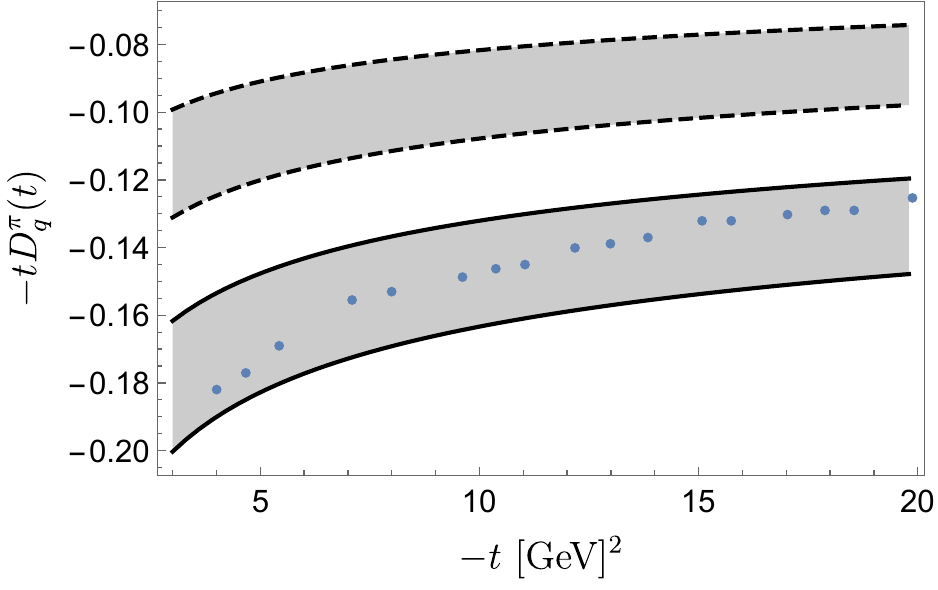}
\includegraphics[scale=0.34]{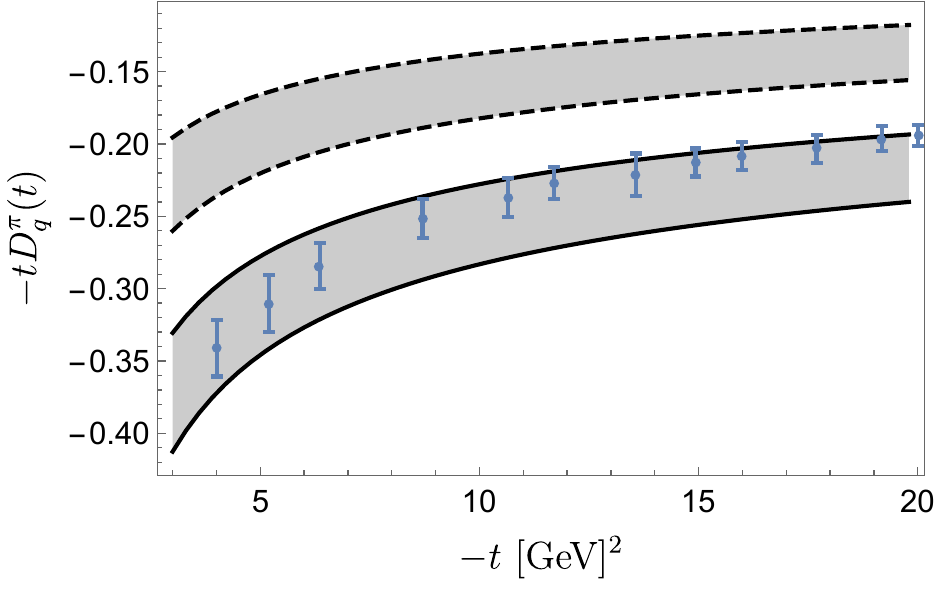}
\caption{\footnotesize  {
Same of Fig. \ref{fig:DA_Ag} for 
the pion GFF $q^2 D^\pi_q(q^2)$ evaluated within Eq. (\ref{Eq:Dql}) \cite{Gao:2022vyh}. 
}}
\label{fig:DA_Dq}
\end{figure*}

In this last section, the GFFs are directly evaluated at high momenta from the  convolution of the meson distribution amplitudes with a
perturbative
kernel   (see, e.g., Refs. \cite{Gao:2022vyh,Tong:2021ctu,Tong:2022zax} for details). 
We report the explicit expressions of the meson GFFs in terms of the meson DAs \cite{Gao:2022vyh} valid for the high  $t$:

\begin{align}
\label{Eq:Agl}
    A^\tau_g(t,\mu_R) &=\int_0^1 \int_0^1 dx dy~\Phi^\tau(x,\mu_R)\Phi^\tau(y,\mu_R) \dfrac{8 \pi \alpha(\mu_R) C_F}{-t} \left[ \dfrac{1}{x(1-x)}+\dfrac{1}{y(1-y)} \right]
    \\
    \label{Eq:Aql}
    A^\tau_q(t,\mu_R) &=\int_0^1 \int_0^1 dx dy~\Phi^\tau(x,\mu_R)\Phi^\tau(y,\mu_R) \dfrac{8 \pi \alpha(\mu_R) C_F}{-t} \left[ \dfrac{x+y+1}{(1-x)(1-y)} \right]
    \\
    \label{Eq:Dql}
    D^\tau_q(t,\mu_R) &=\int_0^1 \int_0^1 dx dy~\Phi^\tau(x,\mu_R)\Phi^\tau(y,\mu_R) \dfrac{8 \pi \alpha(\mu_R) C_F}{-t} \left[ \dfrac{x+y-3}{(1-x)(1-y)} \right],
\end{align}
where: $\mu_R$ is the rinormalization scale  set to be equal to the factorization scale \cite{Gao:2022vyh},  $\Phi^\tau(x,\mu_R) = f_\tau \phi^\tau(x,\mu_R)/(2\sqrt{6})$, being $f_\tau$ the  e $\tau$ decay constant, $\phi^\tau$ is the $\tau$ DA and $C_F=4/3$. 
In Ref. \cite{Gao:2022vyh} the authors suggested three different scenarios for the calculations in order to fix the renomarlization scale: $a)~ \mu_R^2 =-t$, $b)~ \mu_R^2 = -4t$ and $c)~\mu_R^2 =- t/4$.
In Fig. \ref{fig:DA_Ag} we compare the calculations of the gluon GFF $A^\pi_g(t)$, Eq. \eqref{Eq:Agl} for the three values of $\mu_R$, with those extracted from  Ref. \cite{Gao:2022vyh}. We found out that the GSWL2 parametrization is capable to reproduce the lattice outcomes. On the contrary, the lattice data for the GFFs $A^\pi_q(t)$, Eq. \eqref{Eq:Aql}, and $D^\pi_q(t)$, Eq. \eqref{Eq:Dql}, agree with 
the GSWL1 predictions, see Fig. \ref{fig:DA_Aq}-\ref{fig:DA_Dq}, respectively.
It can be inferred that the GSWL1 set is more appropriate for describing quark contribution,  while the GSWL2 set  is capable of capturing the behavior of the gluon distribution.

\begin{figure*}
\includegraphics[scale=0.5]{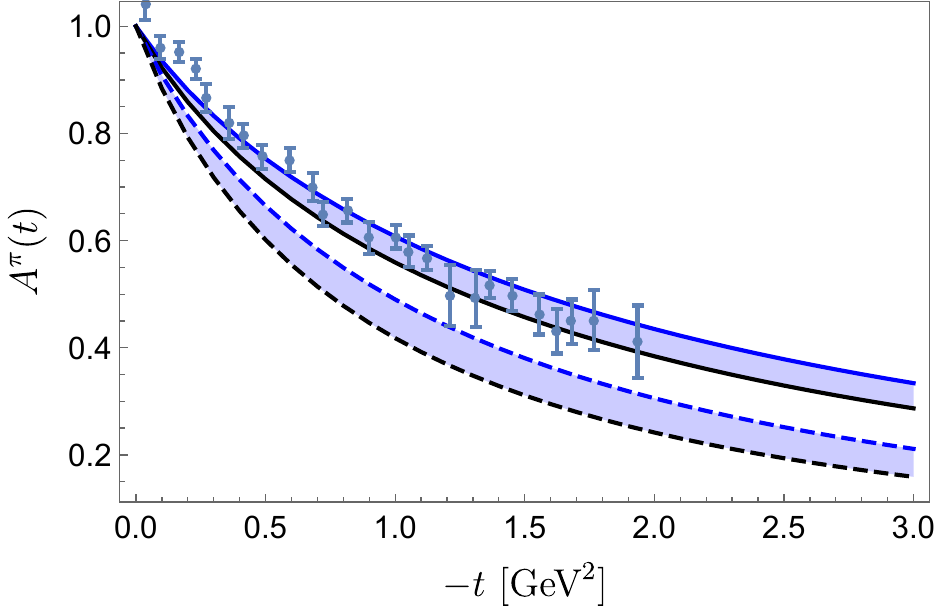}
\hskip 0.5cm 
\includegraphics[scale=0.5]{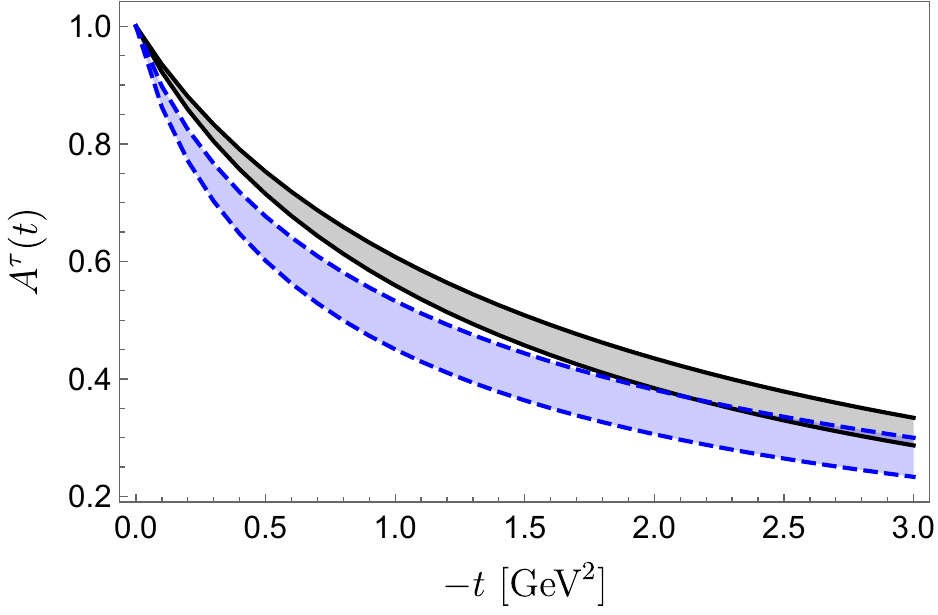}
\caption{\footnotesize  {
The GFF $A^\tau(t)$  Eq. \eqref{Eq:A_LF}.
Left panel: for the pion.  Full lines for the GSWL1 and dashed-line for the GSWL2 parametrizations, respectively. Lattice data from Ref. \cite{Hackett:2023nkr}. Right panel: comparison between the $\tau=\eta$ GFF (dashed lines) with that of the pion (full line), evaluated with the GSLW1  set. Bands
stand for the error in $\alpha$. 
}}
\label{fig:A}
\end{figure*}

\begin{figure*}
\includegraphics[scale=0.5]{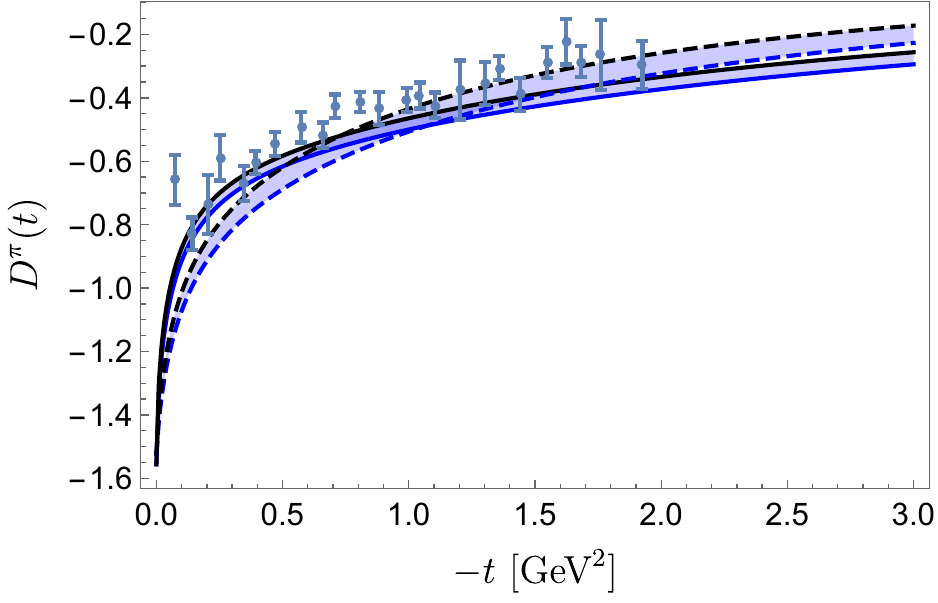}
\hskip 0.5cm 
\includegraphics[scale=0.5]{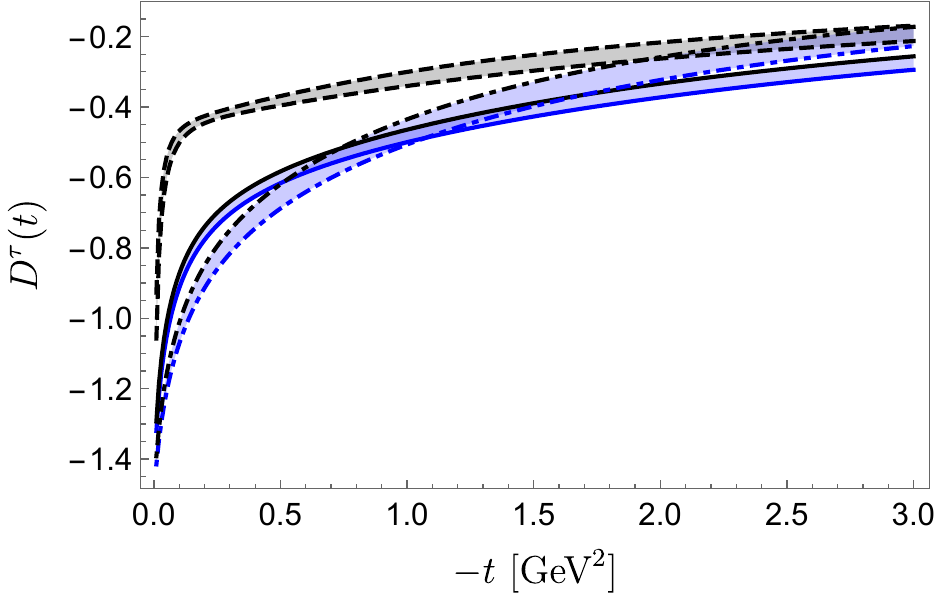}
\caption{\footnotesize  { 
The GFF $D^\pi(t)$.
Left panel: the pion.  Full lines for the GSWL1 and dashed-line for the GSWL2 parametrizations, respectively. Lattice data from Ref. \cite{Hackett:2023nkr}. Right panel: comparison between the $\eta$ GFF (dashed lines)  with that of the pion (full line for the GSLW1 set and dot-dashed lines for the GSLW2 one). Bands
stand for the error in $\alpha$.  
}}
\label{fig:D}
\end{figure*}

\begin{figure*}
\includegraphics[scale=0.34]{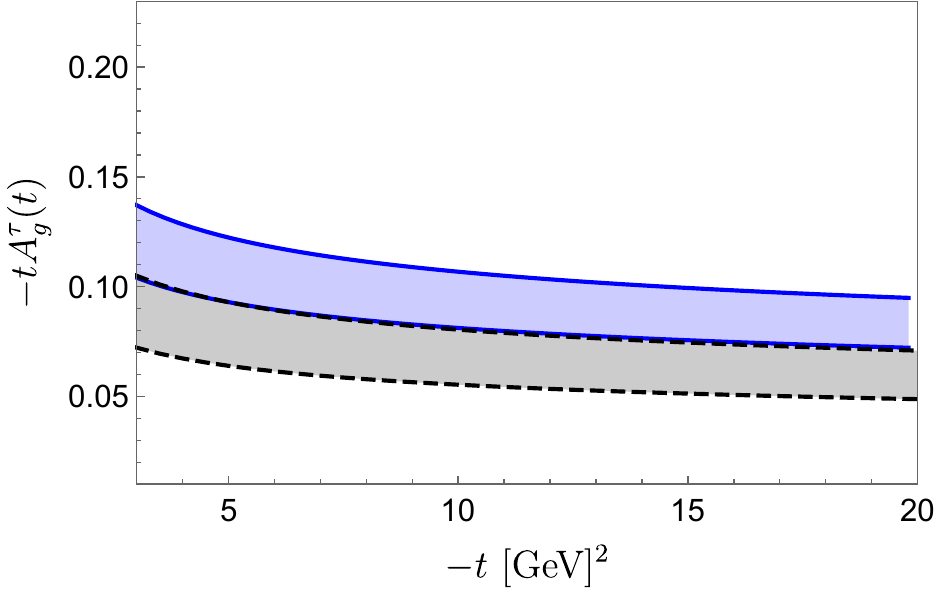}
\includegraphics[scale=0.34]{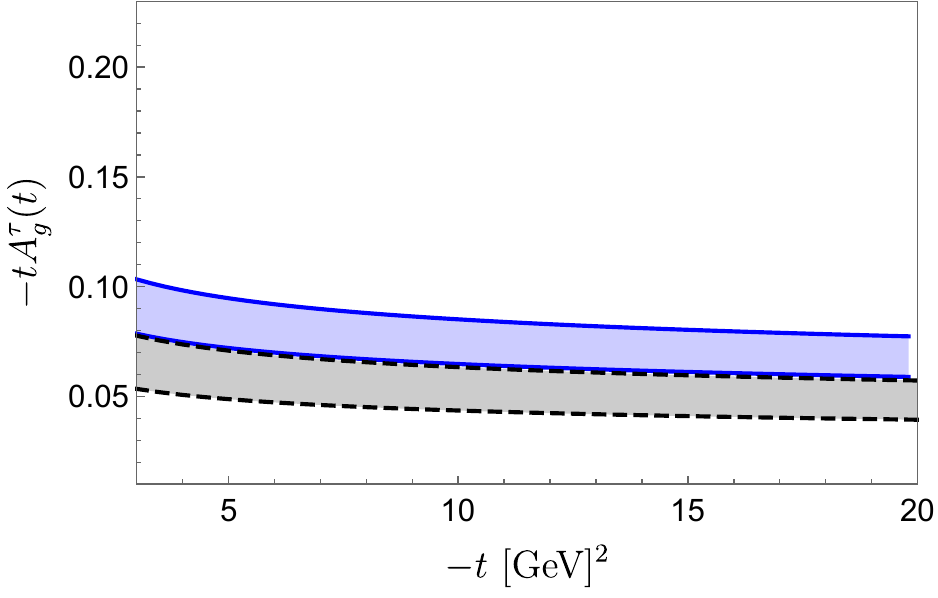}
\includegraphics[scale=0.34]{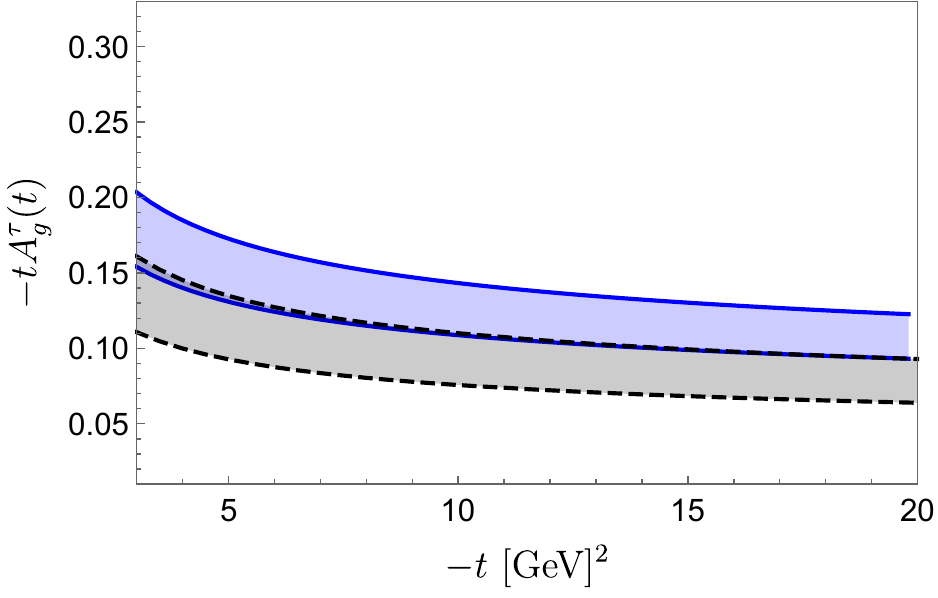}
\caption{\footnotesize  {
Comparison between the pion GFF $A^\pi_g$ (full lines) with that of the $\eta$ (dashed lines).
 Use has been made of the GSWL2 parametrization for the pion distribution.
}}
\label{fig:pi_eta_Ag}
\end{figure*}

\begin{figure*}
\includegraphics[scale=0.34]{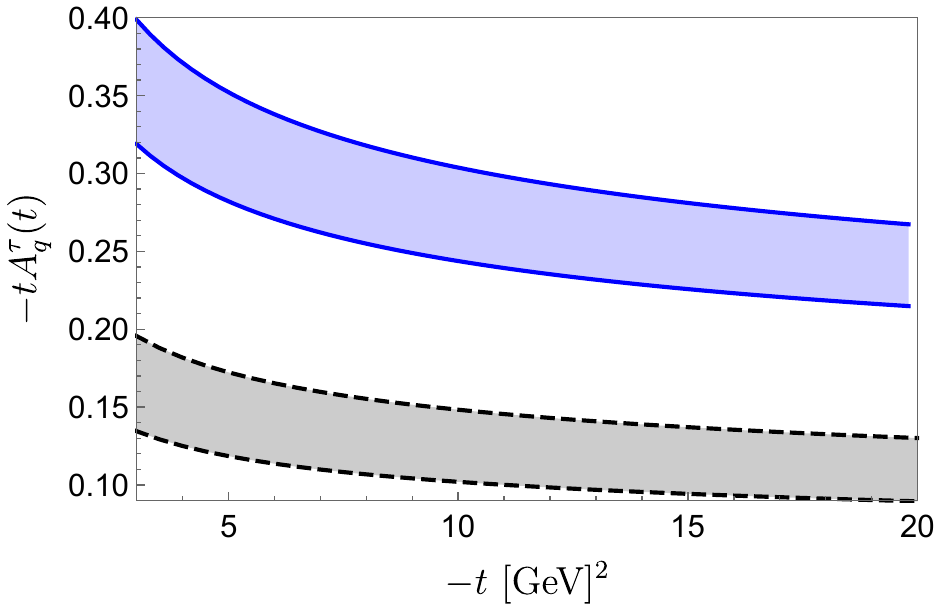}
\includegraphics[scale=0.34]{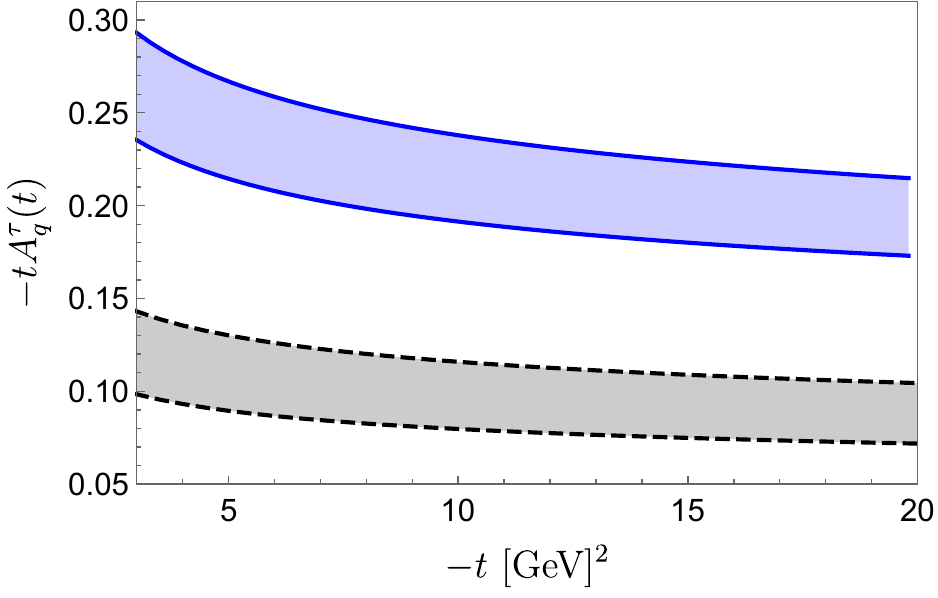}
\includegraphics[scale=0.34]{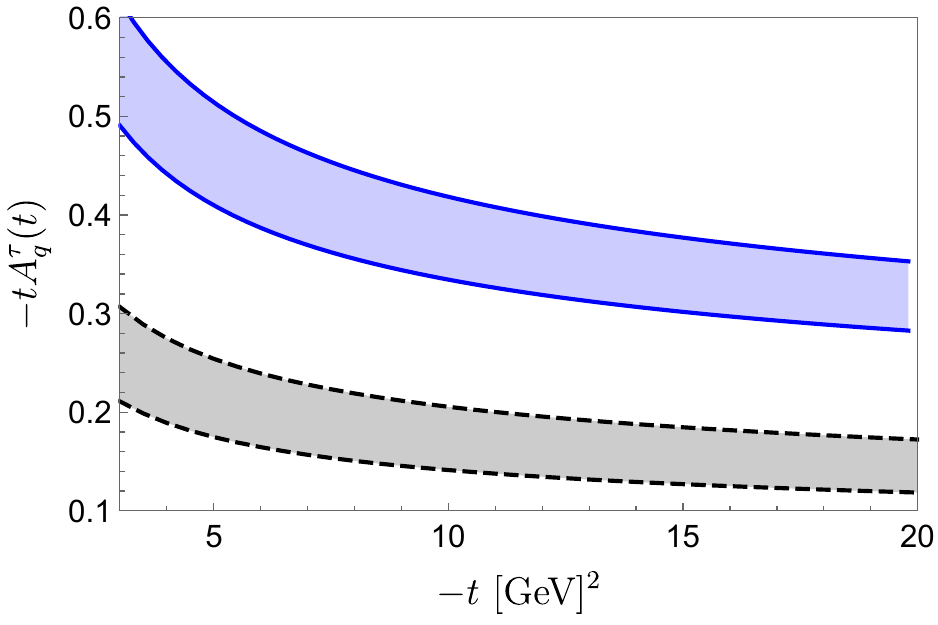}
\caption{\footnotesize  {
Same of Fig. \ref{fig:pi_eta_Ag} but for
$A^\pi_q(t)$. 
}}
\label{fig:pi_eta_Aq}
\end{figure*}

\begin{figure*}
\includegraphics[scale=0.34]{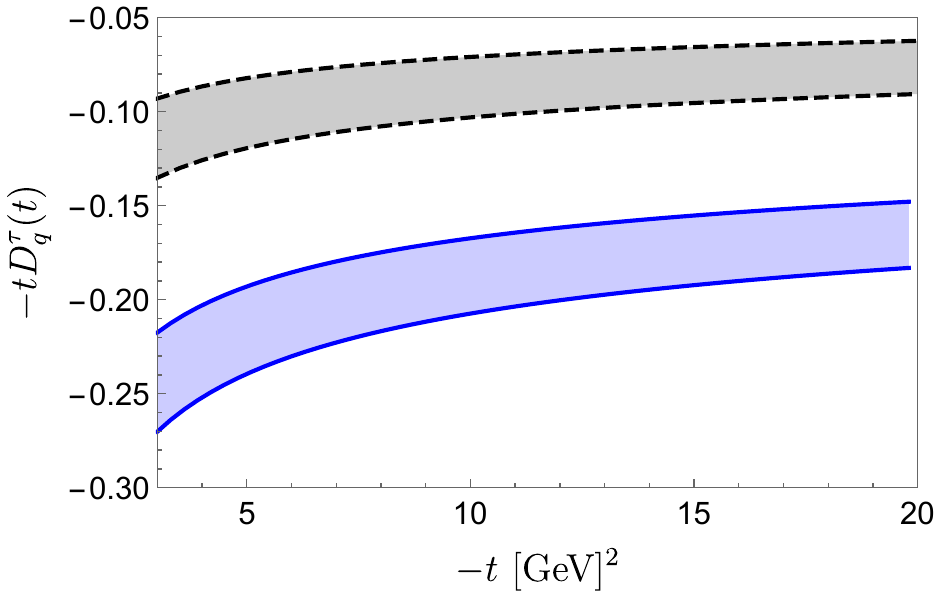}
\includegraphics[scale=0.34]{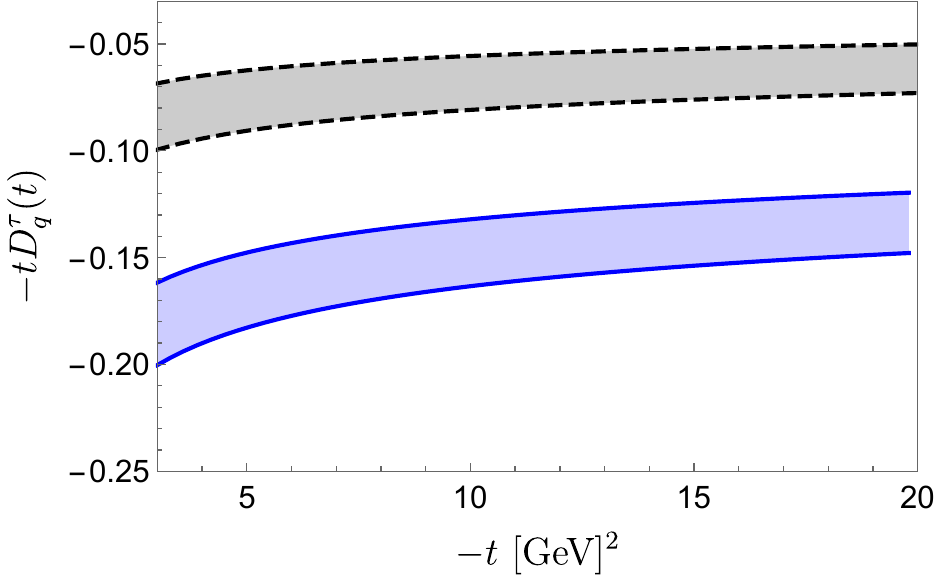}
\includegraphics[scale=0.34]{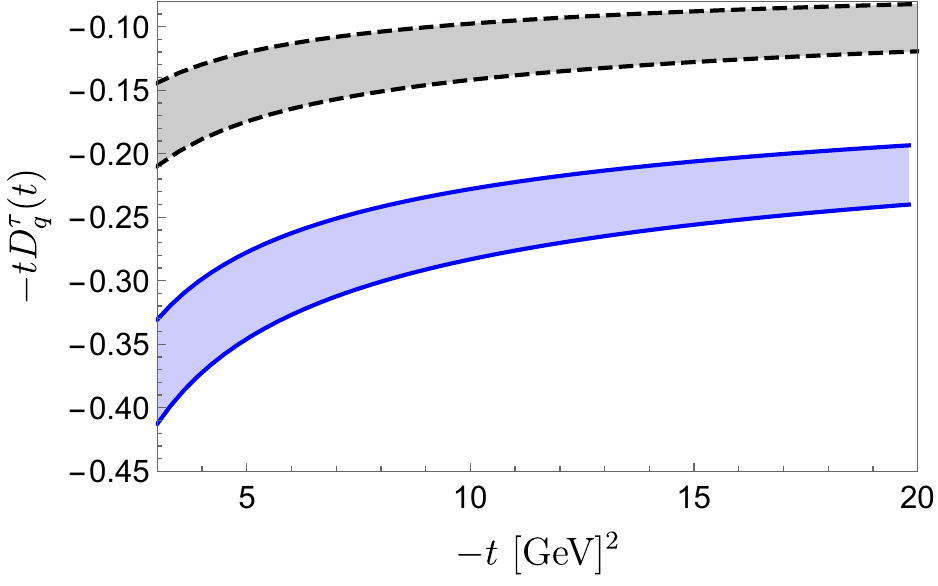}
\caption{\footnotesize  {Same of Fig. \ref{fig:pi_eta_Aq} but for$D^\tau_q(t)$. 
}}
\label{fig:pi_eta_Dq}
\end{figure*}

Let us conclude with the first predictions of the $\eta$ GFFs evaluated from the corresponding DA. Results are shown  in comparison with those of the pion. We make use of the GSW parametrization which better reproduces the lattice data of the GFF of a given flavor, as shown before.
In Fig. \ref{fig:pi_eta_Ag} we display the gluon contribution to the GFF $A^\tau_g(t)$. In this case, use has been made of the GSLW2 set of parameters for the pion case. As one can see, the $\eta$ and $\pi$ GFFs can be  close to each others, within the theoretical uncertainty,  in  all three scenarios of different rinormalization scales. On the contrary, in the quark case, for both the GFFs $A^\tau_q(t)$ (cf. Fig. \ref{fig:pi_eta_Aq}) and $D^\tau_q(t)$ (cf. Fig. \ref{fig:pi_eta_Dq}), the pion results are distant from those of the $\eta$. Finally, also in this framework we recover the behaviour  previously observed in Figs. \ref{fig:A}-\ref{fig:D}.
A possible interpretation of these results is that, since the model does not account for Fock states containing non-perturbative gluons, the perturbative gluon distributions of the pion and $\eta$
 meson are close to each others. Conversely, as the underlying dynamics governing these two $q\bar{q}$ states differ, the corresponding quark distributions are distinct.

\begin{table}[h]
\centering
\begin{tabular}{|c|c|c||c|}
\cline{2-4}
\multicolumn{1}{c|}{} & \multicolumn{2}{c||}{$\pi$} & $\eta$ \\
\cline{2-4}
\multicolumn{1}{c|}{} & GSWL1 & GSWL2 &  \\
\hline
$\sqrt{\langle r^2 \rangle_{mat}}$ [fm] & $0.3045 \pm 0.015$ & $0.3707 \pm 0.025$ & $0.436 \pm 0.041$ \\
\hline
$\sqrt{\langle r^2 \rangle_{mech}}$ [fm] & $0.3053 \pm 0.014$ & $0.47 \pm 0.03$ & $0.47 \pm 0.03$ \\
\hline
\end{tabular}
\caption{Pion and $\eta$ meson mechanical and matter mean radii, respectively. The theoretical error is due to the variation of the parameter $\alpha$. }
\label{Tab:radii}
\end{table}

\newpage

\begin{figure*}
\centering
\includegraphics[scale=0.4]{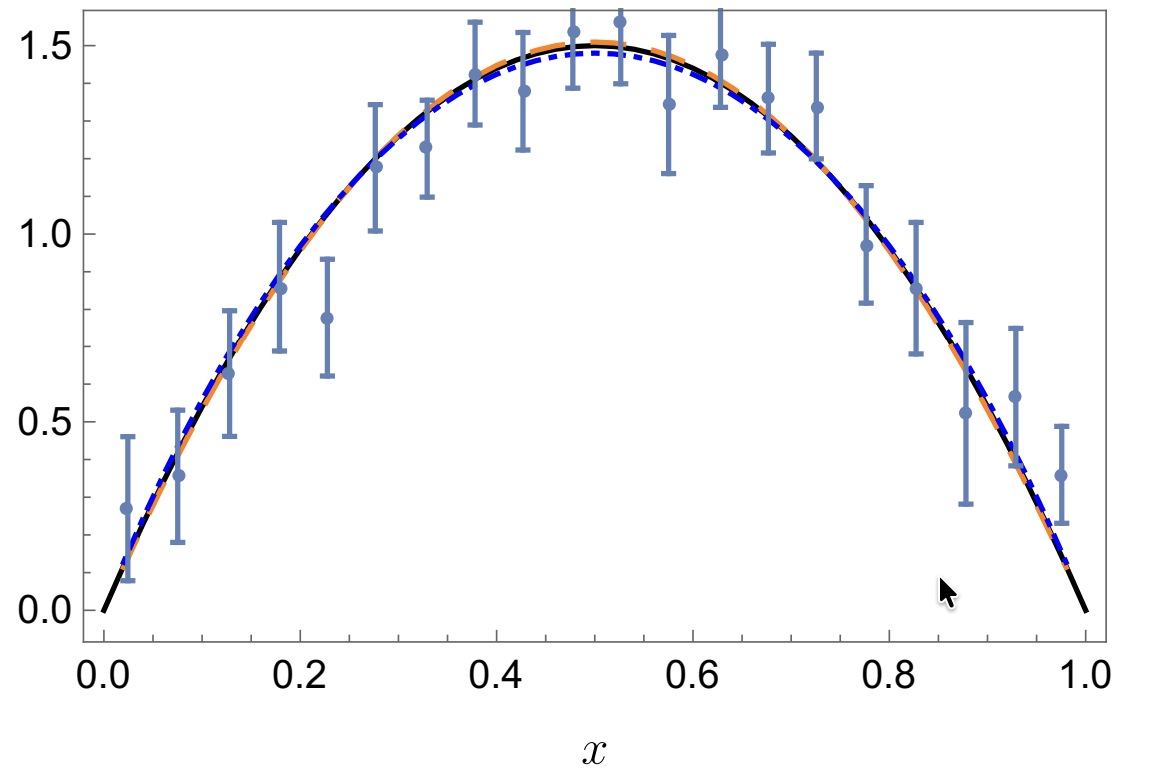}
\caption{\footnotesize  {The pion DA $\phi^\pi(x, \mu_R = 3.16~[\mbox{GeV}])$. Data from Ref. \cite{E791:2000xcx}. Full line for the pQCD predction \cite{Lepage:1980fj,Brodsky:1984xk,Radyushkin:1977gp}. Dashed line for the GSLW2 parametrization and dot-dashed line for GSLW1 one. Figure from Ref. \cite{Rinaldi:2022dyh}.
}}
\label{fig:DA}
\end{figure*}

\newpage

\section{Conclusions}
In this study we have evaluated the pion and the $\eta$ meson gravitational form factors, parametrazing the  Lorentz structure of the energy momentum tensor matrix element and encoding fundamental information on the mechanical properties of hadrons, 
from moments of GPDs  at non-zero skewness. 
In order to estimate these quantities, use has been made of the  meson  wave-functions obtained from the graviton soft-wall model. We remind that the original  GSW approach predicted the $\eta$ meson as the lightest pseudo-scalar. Therefore, in order to describe the pion, the longitudinal dynamics has been taken into account to model  the chiral-symmetry breaking mechanism.

Within this framework, the $H$ GPD of both  mesons have been calculated in its full dependence on transverse and longitudinal momentum transfert.
Then, thanks to the polynomial property of GPDs, both the GFFs $A(t)$ and $D(t)$ have been evaluated. 
The predictions for these quantities are found to be in good agreement with all lattice results.
From the $D(t)$ GFF, the pressures and shear-force distributions have been also calculated  fulfilling  the  von Laue stability condition. Also the normal and tangent force distributions have been studied and we addressed how the normal force is positive definite, as expected for a stable system. On the contrary, the tangent distribution has a node reflecting how the stretching contribution is dominant 
at small distances while the tangent force is squeezing   elsewhere. 
All these quantities have been  studied for the $\eta$ meson in order to compare its mechanical properties with those of the pion. The model predicts that  the $A(t)$ GFF of the $\eta$ is smaller w.r.t. that of he pion, at variance of   the GFF $D(t)$. Therefore, the matter mean radius of $\eta$ is bigger w.r.t.  that of the pion.
This feature is also observed for the mechanical mean radius obtained with the GSLW1 wave-function parametrization, whereas the two results coincide for the GSLW2 set.
In addition, both  stretching and squeezing contributions to the tangent force of the pion is more intense w.r.t. that
of the $\eta$. Finally, the GSW model has been used to evaluate the GFFs in the perturbative region of the momentum transfert  and results are in good agreement  with the lattice calculations also in this case.
In view of future lattice analyses, also the $\eta$ GFFs have been evaluated within the same procedure. 
In closing, the GSW model is capable to reproduce the present lattice data for the pion GFFs, also in the perturbative regime without the addition of extra parameters, hence confirming the relevant predictive power of the model itself.

\newpage

\bibliographystyle{unsrt}

\bibliography{GSW.bib}

\end{document}